\documentclass[10pt,a4paper,twocolumn ]{article} 
\usepackage[T1]{fontenc}
\usepackage[utf8]{inputenc}
\usepackage{calc}
\usepackage{indentfirst}
\usepackage{fancyhdr}
\usepackage{graphicx,epstopdf}
\usepackage{lastpage}
\usepackage{ifthen}
\usepackage{lineno}
\usepackage{float}
\usepackage{amsmath}
\usepackage{setspace}
\usepackage{enumitem}
\usepackage{mathpazo}
\usepackage{booktabs} 
\usepackage[largestsep]{titlesec}
\usepackage{etoolbox} 
\usepackage{tabto} 
\usepackage[table]{xcolor} 
\usepackage{soul} 

\usepackage{multirow}
\usepackage{microtype} 
\usepackage{tikz} 
\usepackage{totcount} 
\usepackage{authblk}
\title{Recommender Systems for the Internet of Things: A Survey}
\author[1]{May Altulyan}
\author[1]{Lina Yao}
\author[2]{Xianzhi Wang}
\author[1]{Chaoran Huang}
\author[1]{\\Salil S Kanhere}
\author[3]{Quan Z Sheng}

\affil[1]{The University of New South Wales, Australia}
\affil[2]{University of Technology, Sydney, Australia}
\affil[3]{Macquarie University, Australia}
\date{}
\begin{document}
\maketitle
\abstract{Recommendation represents a vital stage in developing and promoting the benefits of the Internet of Things (IoT). Traditional recommender systems fail to exploit ever-growing, dynamic, and heterogeneous IoT data. This paper presents a comprehensive review of the state-of-the-art recommender systems, as well as related techniques and application in the vibrant field of IoT. We discuss several limitations of applying recommendation systems to IoT and propose a reference framework for comparing existing studies to guide future research and practices.}

\section{Introduction}\label{intro}
Recent advances in identification technologies, such as wireless sensor networks, RFID, and nanotechnology, have empowered a multitude of physical things with lifted computing ability \cite{sheng2017managing}. Everyday things have become recognizable, addressable, and controllable over the Internet. The potential for seamlessly integrating the physical and cyberspace has created enormous business opportunities. However, finding an effective mechanism for searching and recommending things remains a significant challenge. Recommender systems present a critical stage in promoting and investigating the advantages of IoT. They generally include procedures that facilitate consumer choices based on their preferences. Given the huge amount of information that is available through IoT, users are likely to receive numerous recommendations for services or products. Knowledge of user preferences is essential for the building of any recommender system. The most important feature of such a recommender system for IoT is its ability to exploit knowledge of human behaviour and other IoT data in order to produce accurate recommendations. With billions of IoT resources connected to and accessible on the Internet, a key question is: How can IoT data be effectively exploited as a source to build recommendation systems?

With the development of new approaches and improvements to conventional recommendation approaches and techniques, numerous recommender systems for IoT (RSIoT) have been developed and implemented in a variety of domains, such as smart homes, smart health, smart car parks, and smart tourism. However, RSIoT still has some new challenges. These are more complex than the conventional recommender approaches for three main reasons \cite{yao2019rec}:
\begin{itemize}
\item	Dealing with and analysing a massive amount of hugely heterogeneous data requires both comprehensive analysis and identification to conduct accurate recommendations.
\item	Exploiting rich contextual information is needed to provide recommendations that match the user preferences, while challenges like resource constraints could obstruct this process.
\item	IoT data are required to pass through several layers for extensive processing during their entire life cycle in order to be inferred and produce recommendations to the end-user. Accordingly, providing security for RSIoT usually requires additional complex security layers, which could affect the system’s overall performance. 
\end{itemize}
The last few years have witnessed some tremendous studies on recommendation systems. However, they have focused on either conventional recommendation techniques, (e.g. collaborative filtering (CF) \cite{bobadilla2013recommender,nagarnaik2015survey,sharma2013survey,sridevi2016survey}) or their applications (e.g. social recommender systems \cite{dou2016survey}). Although there are some reviews and outlooks for general recommendation systems \cite{singh2017survey}, no previous research has presented a comprehensive analysis of RSIoT. For example, Burke et al. \cite{burke2002hybrid} reviewed a landscape of hybrid recommender systems and compared them with traditional recommendation approaches. The work in \cite{lu2015recommender}  examined 177 papers on recommendation systems, classifying them into two types and describing their techniques and applications. The authors in \cite{ashley2015survey,abdulkarem2019context,yao2018collaborative}  focused on context-aware recommender systems and illustrate their algorithms. Evaluating a recommender system is usually considered to be a major challenge. In \cite{silveira2019good}, the authors summarized the metrics to evaluate a recommender system and classify them based on their user dependency characteristics. Singh et al. \cite{singh2017survey} presented the three generations of recommender systems and discuss their similarity measures and evaluation metrics. Despite the significance of the RSIoT for both researchers and real-world developers, no previous research has, to the best of our knowledge, comprehensively reviewed recommendations systems in the IoT environment. The main contributions of our paper are summarized below:
\begin{itemize}
\item	We discuss the challenges of using IoT for recommendation systems.
\item	We conduct a comprehensive review of recommendation techniques for IoT and discuss related studies.
\item	We review the development of applications for RSIoT in a variety of domains.
\item	We provide an analysis and its findings for the state of the art.
\item	We provide a reference framework to compare the existing studies and to guide future research and practices.

\end{itemize}
The paper is organized as follows: Section~\ref{sec:Overview} introduces the preliminaries for RSIoT; we also discuss the various limitations associated with the use of RSIoT. Section ~\ref{sec:Recommender Systems} analyses the growth of RSIoT techniques: State of Art. Section ~\ref{sec:Applications}  includes applications of RSIoTs. Section ~\ref{sec:Analysis} provides the analysis and findings. Section ~\ref{sec:Unified} provides a unified recommender system for the IoT framework and future research directions, and Section ~\ref{sec:Conclusion} concludes the paper.

 \section{Overview of Recommender Systems and Internet of Things}
 \label{sec:Overview}
    
This section explains the basic terminology and concepts regarding the recommender system and IoT. We also discuss the limitations of introducing an RSIoT.
\subsection{Recommender System(RS)}

RS proactively recommends items that users may prefer. It has evolved through three main generations from RS for E-commerce, context-and social-aware RSs, and RSs that seek to handle IoT data \cite{singh2017survey}. Several approaches are used to build RSs; however, the conventional approach comprises three main categories: collaborative filtering, content-based, and hybrid RSs. Collaborative filtering recommends items for a particular user based on the ratings of previous users. Content-based methods recommend items from the same category as the items that the user has targeted before. The hybrid approach combines two or more recommendation methods. There are three major requirements for an effective service recommendation as discussed below.
\begin{itemize}
\item \textbf{\textit{Accuracy}.} This is a critical metric in any recommendation system as it ensures the appositive experience of users. An RS is defined as accurate if it recommends relevant services and a few irrelevant ones, especially when there is a lack of information to support the recommendation.
 \item \textbf{\textit{Productivity}}.One of the critical requirements for RSs is productive ability. Productive recommendations are those that are able to produce without the user’s explicit request. Everything can connect to the Internet from any place and for anyone, especially with IoT. Therefore, rich information is available to decide if this situation needs a recommendation or not, or to define which kinds of recommendation techniques are to be employed. 
 \item \textbf{\em Diversity.}Unlike traditional recommendations, RSs in the IoT environment need to handle heterogeneous and interdependent relationships between entities, data, information, and knowledge. Consequently, it becomes crucial to provide the techniques in order to distinguish between the various kinds of relations and to finally conduct an accurate recommendation
    
 \item \textbf{\textit{Newly deployed services.}} It is critical that newly-deployed services can address the cold start problem, which is considered to be one of the main issues for RSs.

    \end{itemize}

\subsection{Internet of Things (IoT)}
 IoT allows internet-enabled physical things to connect, communicate and exchange data. RFID, wireless sensor networks and embedded objects, known as smart things, form a network that bridges the physical and virtual worlds. Smart mobiles, multimedia appliances, toys, and all kinds of other devices can be embedded with sensors to participate in the network. The IoT can be considered a combination of ubiquitous computing, pervasive computing, and mobile computing. The ultimate goal of IoT is to provide a seamlessly integrated platform through which applications involving either physical things or traditional virtual resources can be developed and integrated through the internet. In other words, we can treat physical things as traditional web resources so as to access and interact with them. The IoT relies on the existing internet standards and architectures, and researchers are seeking ways of reusing and adapting current internet standards, such as TCP/IP, HTTP, and Web services, to physical things. 
However, designing an RS based on IoT is far more complicated than designing traditional RSs because of several drawbacks and limitations that are associated with the use of IoT, as described below.
\begin{itemize}
    \item \textbf{\textit{Big data management}}.IoT devices generate massive amounts of data, so an efficient big data management system is needed to deal with various kinds of data and variable velocity. Providing reliability and scalability for this system is important to ensure that it works with no downtime. 
\item \textbf{\em Trust management.} This challenge arises when the RS deals with a large distributed sensors network. The system should have the ability to defend against malicious nodes by establishing a technique to detect untrustworthy entities. One technique to ensure this trust and reputation among IoT devices is allowing each IoT device to evaluate the trustworthiness of the others \cite{sun2007trust,altulyan2019unified}.
   \item \textbf{\textit{Privacy}}. RSIoTs deal with a huge amount of sensitive data concerning individuals, people, organizations, businesses, and health providers. Authentication, authorization, data encryption and integration, and fine-grained access control are crucial to protect these data \cite{frustaci2018evaluating}.
    \item \textbf{\textit{Security}}. Key security risks can be grouped into three categories: risks connected with physical components, such as counterfeit attacks, false attacks, information tampering and network damage; communication risks, such as DoS and DDoS attacks; and risks associated with applications, such as information disclosure, authentication, illegal human intervention, and unstable platforms \cite{misra2017security}. However, dealing with security issues for RSIoT is more complex than traditional RS because of the heterogeneity of the objects and the large scale of the network. 
\item \textbf{\em Interoperability}. Enabling communications among IoT devices that have different standards and protocols can be a critical issue. Networking protocols should be adapted to eliminate the restrictions between constrained and unconstrained entities in the IoT environment \cite{scuotto2016internet}. 
  \item \textbf{\em Quality of services}. Quality of services is one of the main requirements of any system in IoT that ensures availability, efficiency, scalability, and adaptability. It is crucial for IoT systems to plan effectively, use resources efficiently, and respond to queries immediately and adaptively in a highly dynamic environment \cite{gubbi2013internet}.
     \item \textbf{\em Heterogeneity.} There might be diverse resources in IoT, so significant distinctions among these resources need to be hidden to provide a consistent presence. 

\end{itemize}

\section {Recommender System for the IoT(RSIoT)}
    \label{sec:Recommender Systems}

\begin{figure*}[!ht]
            \centering
            \includegraphics[width = 0.85\linewidth]{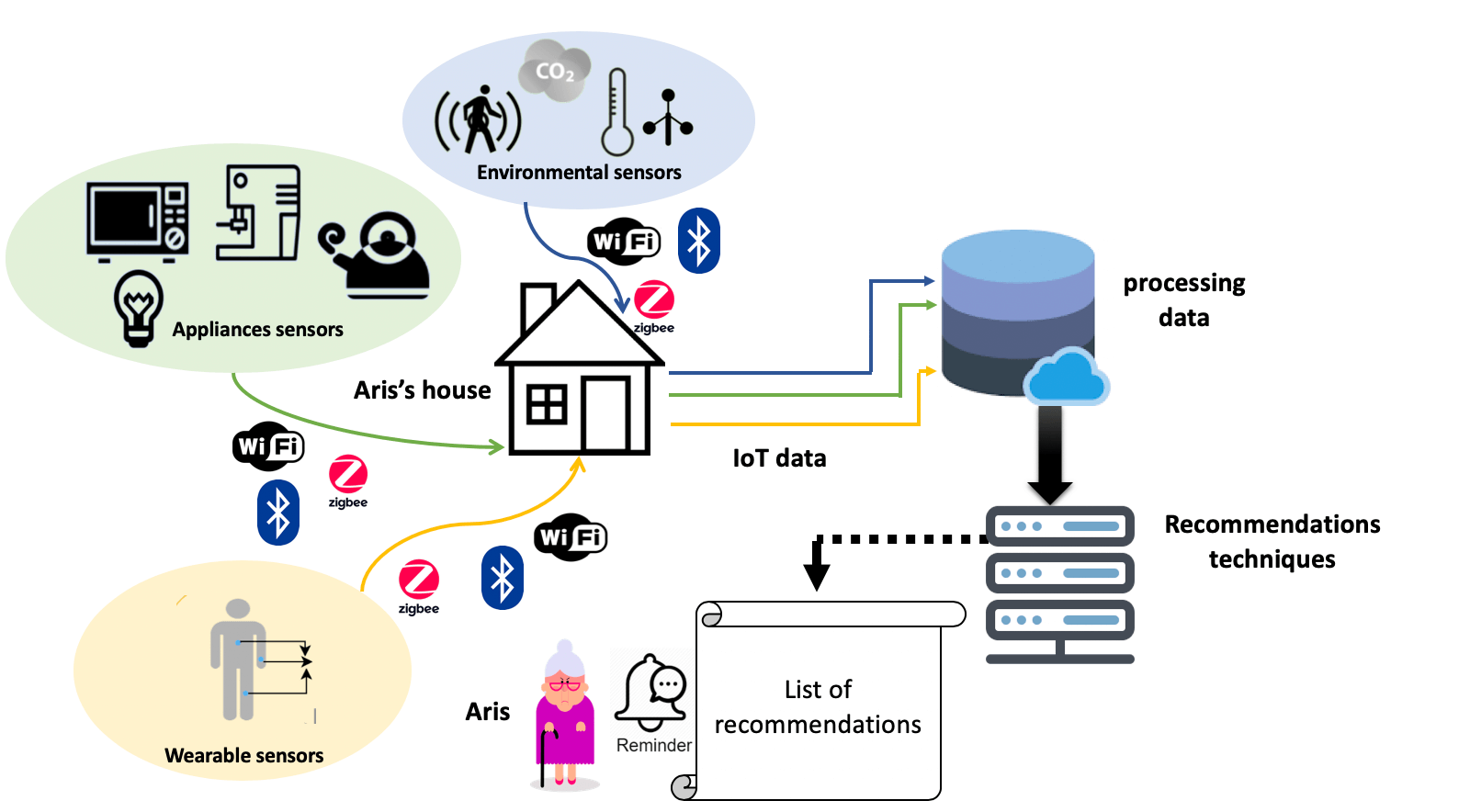}
            \caption{ Motivating scenario of a recommender system for IoT. }
            \label{fig34}
        \end{figure*}

An RSIoT has the attractive property of promoting the advantages that IoT has in supporting individuals, businesses, and society. Similar to a general RS, it facilitates IoT-based systems by proactively delivering things of interest to users, based on their preferences. Meanwhile, it saves the time and cost of using IoT resources in specific situations. The importance of RSIoT can be illustrated by the case of Alice, an 80-year-old woman with dementia (see figure\ref{fig34}), who lives alone in a house and is preparing a cup of coffee in her smart kitchen. Motion sensors monitor her every move and track each coffee-making step. If she pauses for too long, a recommender application will remind her of what to do next. If she tries to prepare a cup of coffee late at night, the system considers the time and recommends she goes back to bed instead. Later that day, Alice’s son accesses the secure application and scans a checklist from the phone in his mother’s house. He finds that his mother has taken her medicine on schedule, slept, eaten regularly, and continued to manage her daily activities on her own.

Before delving into the details of RSIoT, it is vital to understand there are two main differences between traditional RS and RSIoT. These differences are briefly explained below: 
\begin{itemize}
\item Most recommendation processes in traditional RS depend on two main sources: item and user, while RSIoT exploits more than these two sources by including real-time data from sensors in its recommendation process. 
\item  RSIoT considers user preferences as a target, more than managing items, as is the case with traditional RS.

\end{itemize}

            \label{fig20}





\subsection{Aspects of concern}

In this section, we highlight several challenges faced by RSIoT, which constitute the main aspects of concern when designing an RSIoT. 
\begin{itemize}

\item\textbf{Diverse relations}. A common challenge for an RSIoT is that it needs to consider heterogeneous relationships among users, things, data, information, and knowledge due to the poor interoperability between things and data. Therefore, a recommendation technique should be able to conduct accurate recommendations by discovering and leveraging all the different relations. 
\item\textbf{Scalability}. An RS should be able to perform consistently when the whole IoT-based system scales out, and the related data accumulate. An IoT-based system is subject to scaling out in both hardware and software, as more and more individual things are connected into the IoT and continuously contribute to the explosive increase in the number of things. Along with this process, the amount of data may also increase dramatically. For this reason, the recommendation framework should maintain stable performance in terms of providing a reasonable response time and consistent accuracy.
\item\textbf{Dynamicity}. An RSIoT needs to handle three aspects of dynamicity: dynamic discovery of things at the network level, dynamic discovery of user preferences based on their situation, and on-demand real-time recommendation. Besides, the location of things, environmental parameters, and related resources can dynamically change. 
\end{itemize}

\subsection{Recommender Techniques and Systems for the IoT: State of the Art}\label{techniques}

We focus on RSIoT. We describe the techniques to build RSIoT and introduce applications of things of interest to recommendations. To analyse the growth of RSs in the IoT environment, this section reviews the major techniques that are exploited to construct RSIoT, including conventional recommender techniques, context awareness, the social IoT technique, multi-agent algorithms, recommendations with a graph database model, recommendations with machine learning and deep learning techniques, and recommendations with reinforcement learning  techniques.

\subsubsection{Collaborative filtering approach} 

This approach makes recommendations on items for a particular user that are based on the ratings of previous users. It works in three stages: computing the similarities between users; selecting a group of users with the same preferences as the user who needs recommendations, and making a recommendation based on the group ratings. There are two main collaborative filtering (CF) techniques: memory-based CF, and model-based CF. In memory-based CF, user recommendations or predictions of ratings on future items are based on the users’ rating behaviour by using correlations between item to item or user to user. However, the whole training set is used each time to predict the recommendations that particularly affect the performance speed, with a large dataset. This issue could be addressed by pre-calculating the necessary information and then updating them incrementally. 

Model-based CF is more scalable in cases where only the training set is used to build the model. It then uses this model to recommend future ratings. Compared with memory-based CF, model-based CF is considered less accurate because of the large fraction among the item-user values in the training part of the dense dataset \cite{su2009survey}. In the last decade, CF algorithms have become a common approach to building an RS. This involves recommending items based on the history of the user or a group of users with the same interest and maximizing performance, especially when there is enough historical information. Although CF algorithms face an important challenge with new items (cold start problem), they have been implemented successfully in various domains. Some studies have investigated this approach in the IoT context. The following account provides an overview of the use of the CF approach in RSIoT, in which each RS based on this method is explained in detail.  In \cite{yao2014exploring}, the authors proposed \textbf{a unified CF model based on a probabilistic matrix factorization recommender sysetm} that exploits three kinds of relations in order to extract the latent factors among these relations: user-user, thing-user, and thing-thing. A directed weighted graph, probabilistic matrix factorisation, and random walk with restart are all applied to extract these correlations. Such an approach was evaluated by measuring the accuracy. This was based on exploiting correlations among the social network and things with other approaches, based on probabilistic factor analysis. The results showed this approach gives better results than the other methods. 

The author in \cite{asiri2016iot} exploited the CF method to design an \textbf{IoT trust and reputation model} that investigated trust and reputation among IoT nodes. The main phases of this model are: (1) Alpha nodes are determined as being stronger nodes in the IoT network, and are responsible for defining jobs and distributing processing among the nodes in the network; (2) Each node in the network provides ratings for its experiences and consequently, a rating matrix is built. (3) Probabilistic neural networks (PNN) are used to divide nodes in the network that have the same functionality into clusters by calculating the similarity among them; (4) Recommendation weights are calculated, and Quality of Recommendation (QR) is defined as a score of trustworthiness; (5) The sensitivity found in each transaction is defined by using a flag parameter; (6) The trust value between the two nodes is computed; and (7) The nodes are classified into trustworthy and untrustworthy. Here, one of the PNN layers is used to calculate the probability density function (PDF) that defines the probability of the node belonging to either the untrustworthy or trustworthy class. The two main features of this model are: it deals with the cold start problem by predicting the value for each new node, and it classifies the data depending on its sensitivity in enhancing the system’s performance. It is considered that the main goal of RSs is to know user preferences, as this enables them to conduct accurate recommendations. 
 In \cite{chakraverty2018iot}, the CF approach was adapted to address this issue. Here, the authors exploited the weather and location data that was collected by the sensors to provide effective recommendations for the residents of that geographical region. This is called \textbf{the weather and location-aware recommendation system}and the architecture of this RS comprises the following functional blocks: W-Historical Weather Data Processing; S-Historical Sales Data Processing; and R-Making Recommendations. In historical weather data processing, weather data are collected and processed; if a record has missing data, it can be purged either by filling the missing interval by the closest reading or taking the average of previous readings. The Hidden Markov Model is used in this block to address the problem of defining weather conditions during short intervals. In historical sales data processing, two types of historical sales data are collected: temporal data and specific location data. If there are missing data, the system can manage the situation by using a trend model to extract the highest sales during a given time interval at a specific location. In \cite{sawant2017representation, yao2018collaborative} the authors proposed a framework that combines cyber-physical systems and IoT to make recommendations. The system architecture consists of four layers: a selection layer, network layer, services layer, and application layer.

\begin{figure}[!ht]
        \centering
        \includegraphics[width = 0.85\linewidth]{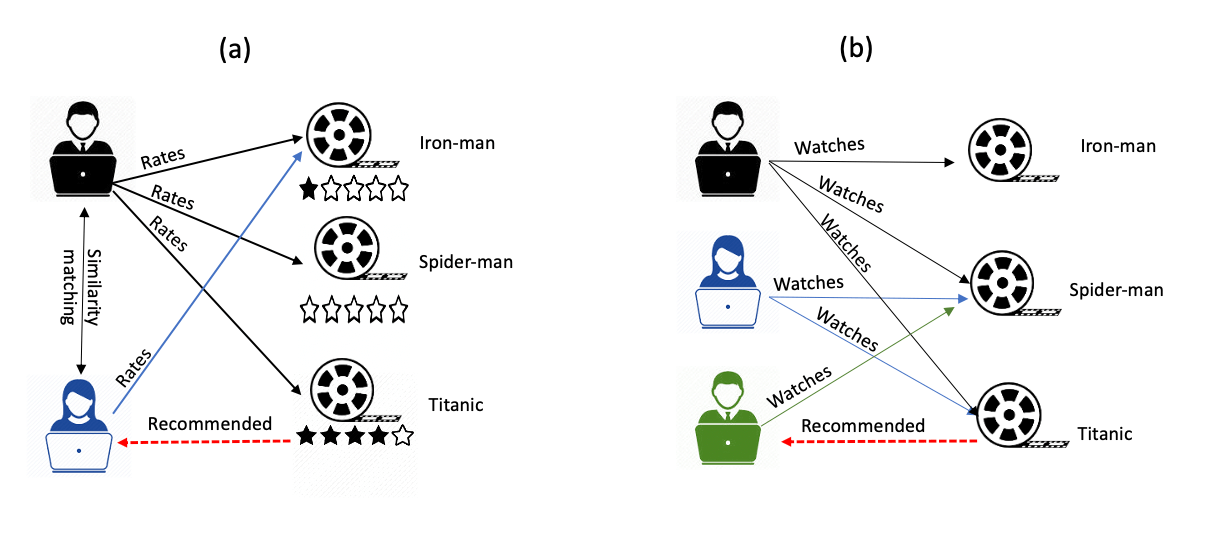}
        \caption{(a)User-based collaborative recommender system,(b)Item-based collaborative recommender system }
        \label{fig2}
    \end{figure}

 The authors discussed the two main approaches to CF to build an RS: a user-based approach and an item-based approach. In the user-based approach, the user plays the leading role; people with the same taste are grouped, and the recommendations for the user are based on evolution items that are rated by the group with the same preferences as the user (see Figure\ref{fig2}(a)). The item-based approach provides recommendations by building neighbourhoods for the items that are preferred by the user (see Fig.\ref{fig2}(b)). The work in \cite{lee2016service} extended the user-based approach to provide group recommendations in an IoT environment by considering the member organization. The member organization can affect the decisions of members of other groups. The system consists of three main parts: (1) constructing the rating matrix from the history accessing services by normalizing the frequency of services access; (2) defining the member organization by using three kinds of similarity metrics: group size-based, common member-based, and member preference; and (3) user-based CF based on group similarity, which is responsible for dealing with the cold start problem, predicting ratings to the new group and producing recommendations. The experiment’s results showed that the member organization-based group similarity (MOGS) metrics outperformed the baseline approach.
Also, in  \cite{mashal2016performance,mashal2016testing}, user-based and object-based are adapted to the RSIoT part in the \textbf{MUL-SWoT model}. The RSIoT is responsible for ranking the services that are received by the third party and filtering them to provide recommendations. The authors use some metrics to evaluate these algorithms; the results show that Object-Based Collaborative Filtering (OBCF) was better than User-Based Collaborative Filtering (UBCF). 

The authors in \cite{yao2015service} exploited matrix factorization to build an RS that provides software recommendations. The system aims to recommend a set of APIs to a specific mashup by using matrix factorization. The basic idea of this system is it exploits two main sources of data to produce recommendations: API profiles and information about their co-invocation in previous mashups. In the first step, the similarity between each pair of API services is calculated by using cosine similarity to evaluate their profile. The matrix factorization maps then mashup into a shared lower dimensional space. Then, based on the factorization results, the probability of API being invoked by each targeting mashup is calculated. The proposed system was tested and evaluated, and the experimental results showed that this approach outperformed the state-of-the-art CF approach. Some studies have sought to make this approach to building an RSIoT more efficient by addressing the major problems of scalability and cold start \cite{zhang2013cruc}. In \cite{nizamkari2017graph}, the authors proposed a system that could recommend the best among several nodes that had the same services. When the recommendations part of the system was built, it was based on the CF approach and optimized it by addressing the two main challenges: the cold start problem and scalability. It tackled the first problem by using a graph-based trust and the second by allowing each node to predict the rating, rather than a central system. The trust and similarity between the nodes were measured, based on the rating and structure of the network. The experiment’s results showed that the combination of Trust as a new influence and Similarity as a traditional influence in CF improved the accuracy and coverage.

In \cite{salis2018anatomy}, the authors proposed an airport fog-to-cloud system, based on the F2C project, which conducts recommendations for both novice and experienced travellers. The main idea of the system is to use the data of traveller’s device to recommend a suitable event, such as a flight’s last call, flight closing, check-in opening, etc. It consists of several entities: Users, who provide all the details about their destination; Sessions, which start when the user contacts any fog node; Position, to define an attribute for both POIs and users. The other entities are Topic, which is chosen by the user to initiate the user preferences; the POI, which represents the user preferences; Promotions, which are based on the POI; and Score, which considers the entities to conduct effective recommendations. To recognize a user’s preferences, the CF adapts. This is based on either the historical actions of the user or similarities to other users. 

For medical related problems, collaborative filtering cannot be adapted directly. Consequently, the authors in \cite{jabeen2019iot} designed a modified version called Advice-based CF to build an RS which provides suggestions for cardiovascular diseases patients. It combines three major parts that each had unique functions: data collection, classification and recommendations. The first part is responsible for collecting data from the patient by using biosensors. The collected data is thereafter cleaned and filtered for extraction. The second part classifies the cardiovascular diseases into 8 classes. The authors proposed four classifiers of machine learning: SVM, Naive Bayes, random forest, multi-layer perceptron. The last and main part is used to produce recommendations for patients in a remote area. 

The common problem with each RS, whether it is conventional or IoT-based, is a lack of information. Some studies attempt to address this problem by adding social network data as an input source \cite{li2019personalization}. For example, authors in \cite{margaris2017exploiting} combined two main algorithms to improve avenue’s recommendations accuracy: a CF-based algorithm using social network data and the QoS-based algorithm. The main operations of the system have the following steps: first, the QoS algorithm calculates the cost for each avenue and then the atmosphere. The CF computes the impact level of the user’s social friends; but, first, the algorithms are worked offline. After the completion of this offline initialization, the algorithms are executed online to conduct the recommendations in parallel. The recommendations of both algorithms are then combined using the WCombSUMi formula \cite{wu2002data}. Finally, the repository data are updated to maintain the consistency of the system. The study \cite{yang2019exploring} also proposed a location-aware POI RS that exploits three kinds of data sources to conduct accurate recommendations: user rating reviews, POI data, and user data. The system has the ability to provide recommendations for the user when he moves to a new region, even if there is no activity history for him. The Matrix factorization approach is applied to the recommendation engine and this gives the system three main features: accuracy, scalability and flexibility. The system is tested using the public dataset (yelp dataset), which gives a high degree of accuracy. 

In \cite{rossi2016towards}, the matrix factorization approach is exploited to present a framework for museum tour recommendations. Although the collaborative filtering approach has been adapted in numerous studies, as we discussed in previous parts, there are potential problems that make it inefficient for RSIoT, as we discuss below. These are:

    \begin{itemize}
      \item    \textbf{Scalability}RSIoT deals with a large amount of data that needs a computation power to conduct the recommendations, as well as fast response to online user requirements. 
        \item \textbf{Cold start problem}particularly when a new user joins the system. There is insufficient information to identify an appropriate group of previous users for rating purposes or when a new sensor is added to the system. 
\item  \textbf{Sparse data} may affect the accuracy of collaborative RSs, especially RSIoT, which receives huge sensory data from multiple channels, as we elaborated in Alice’s scenario. 
    \end{itemize}

\subsection{Content-based approach}

\begin{figure}[!ht]
        \centering
        \includegraphics[width = 0.85\linewidth]{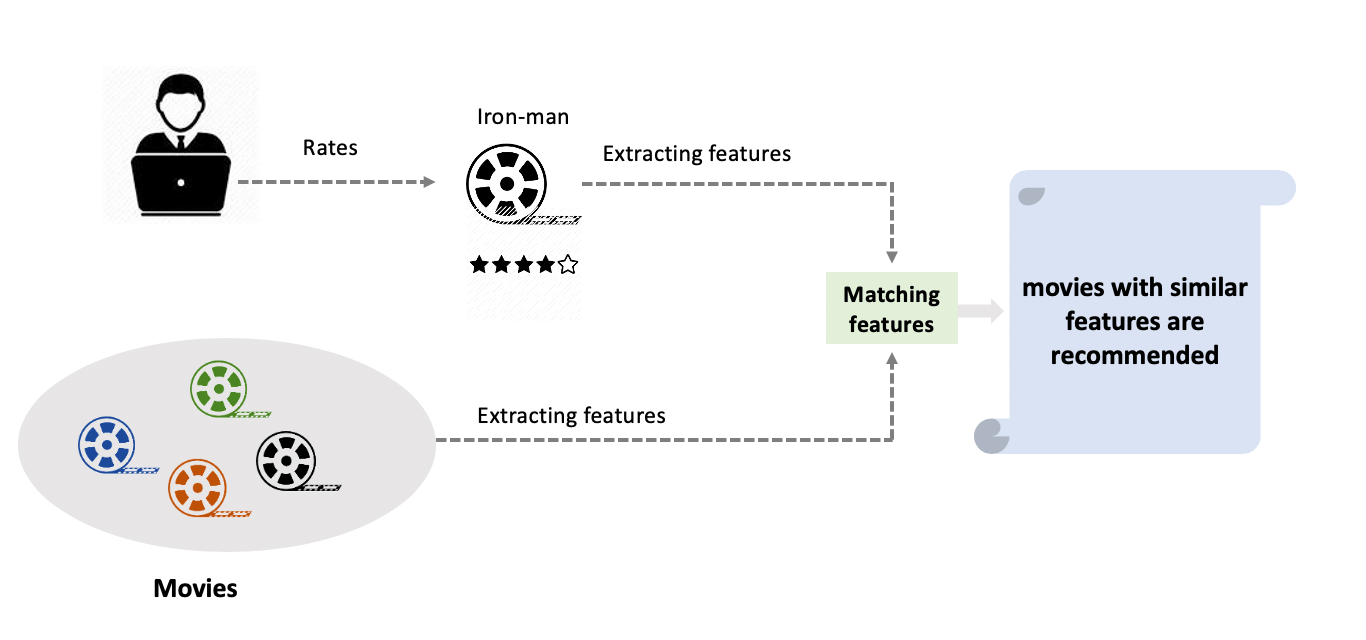}
        \caption{Content-based recommender system .}
        \label{fig4}
    \end{figure}

The content-based (CB) approach recommends items that are similar to the items previously targeted by the user. Instead of relying on ratings, it uses the existing interest history to predict user interest in the target and match the content of similar profiles with the target content (see Figure \ref{fig4}). Therefore, it may suffer from the cold start problem \cite{sawant2017representation}. The CB approach has been used for recommendations in the IoT. 
In \cite{erdeniz2018recommender}, the authors proposed two RSs for their AGILE project, which aims to improve users’ health conditions. As a result, two new apps conduct the patients’ healthcare devices and physical activity plans. The CB approach is used to build the recommender engine for a second app. the idea of the app is to collect the patient’s data through medical sensors that can be worn, and a virtual nurse is illustrated as a case to explain this. Based on these data, the app can recommend a suitable activity plan to the patient. In another illustration, in \cite{koubai2019myrestaurant}, the authors adapted a CB approach to building a recommendation module for their smart restaurant, which aims to provide dish recommendations based on the customers’ tracking history. The Jaccard measure was adapted to measure the similarity between the customers and the dishes to conduct recommendations based on user preferences. The CB approach has also been applied to parking applications. However, based on this technique, the applications will only make recommendations that are based on the similarities among the customers’ parking profile without considering their user preferences. Therefore, accurate recommendations resulted in parking issues. To address this problem, the work in \cite{srisura2019periodical} proposed a periodical recommendation that is conducted during three periods: recommendations at the entry gate, recommendations for the nearest parking zone and recommendations for the new zone, when the recommended parking is taken. The system uses three main dimensions as data sources: user profile, user preferences and nearest zone. The distance matrix is adapted to calculate the distance between the user and a zone. To minimize the computation time, the system can remove unrelated zones from the zone list. The system was also tested using a simulation dataset and the result shows it outperforms other systems that focus on user profile or user preferences only.

    \subsection{Hybrid approach}

The hybrid (HB) approach combines two or more approaches to build an RS. For example, by combining the collaborative and content-based methods, the limitations of each can be addressed. The different ways of merging collaborative and CB approaches into a hybrid RS are as follows \cite{burke2002hybrid}:
\begin{itemize}

\item Building a hybrid RS by implementing content-based and collaborative features separately and mixing their predictions.
\item Combining some content-based features into a collaborative approach. 
\item	 Combining some collaborative features into a CB approach.
\item	 Building a general unifying model that combines collaborative and content-based characteristics. 
\end{itemize}
Significant research efforts have been devoted to exploiting the hybrid approach to improving recommendations and search performance in IoT. In particular, the framework of the recommendations for Digital Signage (DS) in Urban Space \cite{tu2016context} exploited the hybrid approaches to make DS more attractive. The framework has five components: IoT sensing, data pre-processing, recommender engine, DS user interface, and DS data storage. The recommender engine plays the main role in investigating product taxonomy, advertisement taxonomy, condition feature groups, and the DS recommender model of the framework. It can deal with multiple reviewers by switching to various reviewer’s modes. The implemented system’s results were based on a digital signage system deployed in a Taipei shipping mall and showed that both the demographic and context features are crucial factors in providing accurate recommendations. To reduce the latency of recommendations, in \cite{kim2018user}, a user-space customized RS platform system is proposed that will exploit the mobile edge environment. The conceptual architecture combines several components to build an RS that produces high-quality recommendations. The system produces the recommendations by exploiting user feedback data and sensor data that are closed to the user. This system deals with two user scenarios: (1) if this is the first time the user has dealt with the system, the recommendations are based on similarity with other user’s preferences (collaborative filtering); (2) if this is not the first time the user has engaged with the system, the user’s feedback data in the repository is considered in the recommendations (content-based). In \cite{hamlabadi2017framework,saghiri2018framework}, the authors built their RS engine using a hybrid recommendation algorithm which combines CF and CB approaches. 

The RS framework is based on the cognitive process. It consists of three main layers. In the first layer (Requirement layer), the goal of the system is described by using a Cognitive Specification Language (CSL), which it then sends to the next layer. The Things layer observes the environment and provides the recommendations list. The main layer is the cognitive process layer, which contains the cognitive recommender engine that is responsible for receiving the sensory data from the Things layer and executing the recommendation algorithm.

 \subsection{Knowledge-based approach}
        
   This approach recommends items to the user, based on knowledge about the users, items, and their relationships. The Knowledge-based (KB) approach has a functional knowledge base whose performance is based on the identified relationship between a user’s need and a possible recommendation \cite{burke2002hybrid}. It does not depend on user rating or gathering information about specific users to provide recommendations. The KB approach addresses several limitations of other approaches, such as the ramp-up (cold start) problem. Ontology is a formal method of representing knowledge that is central to building RSIoT.

Creating and evaluating ontologies are complex and time-consuming processes, but several studies have used them to build RS s in IoT. For example, the authors in \cite{hwang2016data} proposed a method for generating automatic rules and recommending the best rule, which requires no complicated configuration or extra efforts at the user end. Also, the user has the opportunity to add new rules for newly connected devices. To increase performance, the authors further applied two steps with three ontology models and collected open web data, related to rules by a data pipeline. First, three ontology models were created: (1) Things (knowledge base), which provides all information for things; (2) Context (knowledge base), which provides contextual information about people, environment and things; and (3) Functionality (knowledge base), which links the functionality of context and things. Second, data about the automatic tasks were gathered from the web (web crawler) to be stored in the IFTTT crawled data. Then the collected data from the web were mapped with the data of the three ontology models in order to produce the home rules. The user can add new automatic tasks when a new device is added (sensor) by recommending some tasks to the system. However, adding new automatic tasks might conflict with current tasks. For this reason, the authors defined four potential conflict situations to be solved. In \cite{kumar2017smart}, a conceptual framework called the web of objects (WoO) is proposed that provides smart spaces services recommendations by using semantic ontology. It combines three main layers: The virtual objects (VO) layer is used to represent real-world objects, a composite virtual objects (CVO) layer that accumulates VOs and their information, and the services layer, which is responsible for providing recommendations based on the results of the previous layers. Also, the authors in \cite{varfolomeyev2015smart} adapted ontology formats for smart space technology to implement an RS for historical tourism. Also, for smart health applications, ontology’s main role is to supervise chronic patients and to provide long-term care \cite{di2017fuzzy}. However, a classic ontology cannot define the membership value of risk and uncertain factors precisely; thus, the system provides inefficient results. To address this problem, \cite{ali2018type} adapted a fuzzy ontology to build a system which monitors diabetes patients and recommends specific food and drugs. It combines 4 kinds of fuzzy ontology: patient ontology which contains the history of a diabetic patient, sensor ontology, which includes the data from sensors, knowledge and rules-based ontology, which consists of the rules that help to define the patients’ conditions, and drugs and food recommendation ontology to provide recommendation services. The experiment’s results show that the performance of the proposed system outpaces that of classical ontology. 
        
    \subsection{Context-aware Recommendation }
    Context-aware computing is well known in computer science research. In \cite{dey2001conceptual}, context is defined as "any information that can be used to characterize the situation of entities (i.e. whether a person, place, or object) that are considered relevant to the interaction between a user and an application, including the user and the application themselves. Context is typically the location, identity, and state of people, groups, and computational and physical objects". 

The application of context-recommender systems in the IoT environment faces several challenges, such as context discovery, security, privacy issues, and sharing context, but some researchers have used context awareness in their frameworks to build RSIoTs. \textbf{Proactive multi-type context-aware recommender system} \cite{salman2017model} exploited the user context to define when the recommendations must be pushed to the user, and which kinds of recommendations are to be pushed. The proposed approach is demonstrated in three domains: gas stations, restaurants, and attractions. The authors built the system in two main phases: identifying the situation of the user and the type of recommendations phase, and the item recommender phase. In the first phase, a context-aware management system (CAMS) performs the following tasks: (1) data acquisition, where data about IoT devices, such as sensors, GPS, etc. are collected; (2) data modelling to convert the raw data into understandable form; (3) context reasoning to deduce new knowledge based on the results of the previous stages – the neural network is applied at this stage to generate several scores for the three types of recommendations; and (4) scoring algorithm to define which types of recommendations will be moved to the next phase. In the second phase, a CF approach is used to recommend the items for the user. The proposed framework was adopted as an application, and then tested and evaluated with good results. 
The authors in \cite{hassani2018querying} proposed a \textbf{Context-as-a-Service (CoaaS) recommender system} that uses an IoT context service to enable applications to provide and consume contextual information seamlessly without the need for manual integration among IoT devices. There two core components are context service description language (CSDL) and context service matchmaking (CSM). CSDL enables developers to describe their services regarding semantic signature and contextual behaviour in standard language. CSM is responsible for matching between context request and context services. The CoaaS framework was tested by adapting it to produce smart car park recommendations via an application on a smartphone; this is explained in detail in section \ref{sub:smart car park applications}. The    \textbf{EW4 model} \cite{yuan2015and} exploited contextual information and the mobility of the user’s tweets to provide accurate recommendations. The framework combines both offline and online components. The author has proposed an algorithm using a generative process to model the offline components, which is based on four aspects: who (User), where (Geo), when (Time) and what (Words). Several applications based on this model could be built, such as location recommendations, use prediction and location prediction. The EW4 model’s performance for some tasks was evaluated using two datasets. 

The authors in \cite{yavari2016contextualised} proposed an RS that could provide accurate recommendations by considering contextualization with IoT data. They defined two operations for contextualization: contextual filter and contextual aggregation. In the contextual filter step, data on IoT devices and services are filtered, based on the current context. Contextual aggregation involves combining the filtered data, based on both contextual similarities and relevance. The authors adapted the framework to design a smart parking application. Zhou et al. \cite{zhou2017social} proposed a model which could improve recommendations accuracy by exploiting context-awareness. The model consists of three main components: firstly, the server, which is responsible for providing services recommendations to users; secondly, the user, who provides the context to the server in order to get recommendations; and thirdly, the services provided by the server. The authors designed a Hierarchical Social Contextual Tree (HSCT) algorithm that uses three sources to provide accurate recommendations: service usage frequency, contextual bandit feedback, and social intimacy. The experiment’s result showed that the system outperforms other algorithms, especially in two dimensions, namely scalability, and the cold start problem. Also, in \cite{kaur2019context} the context is adapted as a source to build a recommender engine for recipe recommendations in a smart kitchen. The recommender engine consists of four main components: food item taxonomy, which represents the food items and their attributes in the refrigerator and cupboard; recipe taxonomy, which includes all the recipes and their ingredients; conditional groups, which combines environment and context features; and the recommender model, which utilizes the historical data to conduct recommendations. Based on the previous components, the authors described a motivation scenario to explain the importance of the proposed system. Also, context awareness was employed in building an RS for smart healthcare \cite{casino2018smart}. In the same vein, the authors \cite{hong2017social} proposed an RS that provides social recommendations for cultural heritage. The architecture consists of three main parts: socializing, social-based recommendations and context-based recommendations. The system can deal with the sparsity issue, where the recommendations are conducted, even when the new user uses the system.

    \subsection{Social IoT-based Recommendation Techniques} 
    The concept of social networking is applied to IoT to create a social relationship among things. The authors in \cite{holmquist2001smart} introduced the concept of socialization between things, which focuses on finding solutions to allow smart wireless devices to establish temporary relationships. The main idea behind SIoT is that social objects can select, discover and compose services. Five main kinds of relationships are produced, based on the object’s activities and features, which are explained in detail in \cite{atzori2012social}.These relationships depend on the object’s activities and the features that are to be created and updated. Moreover, they require some modules to utilize these relationships, such as a service discovery module and relationship management module \cite{saleem2016exploitation} SIoT enables IoT applications to exploit IoT services efficiently. Hence the RS can exploit the data that are gathered from various applications. For example, the authors in \cite{saleem2016exploitation} proposed a framework that could exploit the SIoT network to produce services recommendations. The SIoT network includes the profiles of all the IoT applications. The data profiles can be used to provide service recommendations or could be exploited by other IoT applications to search for similar conditions that have been solved in the past. The proposed framework has three main layers: the SIoT perception layer, which collects the data of all IoT devices and their relationships, which are extracted by using SIoT technique; the network layer, which connects the previous layer with the upper layer by using several communications protocols; and the interoperability layer, which enables all IoT applications to use the IoT data. The authors provide an example of how their framework could be adapted for various applications. This would be done by exploiting the availability of social correlations among things and between users and things to provide services recommendations.

However, facilitating access to quality services and trustworthy devices in a SIoT environment has become a critical issue. To address this problem, a scheme of access service recommendation \cite{chen2016scheme} is proposed that provides recommendations about trustworthy nodes. Trustworthiness is evaluated, based on three parameters: a feedback-based reputation system, social relationship and an energy awareness mechanism. Recommendations are subject to an unreliable wireless environment and limited battery capacity in smartphone applications. In \cite{ren2018recommender}, the mobile IoT is exploited to build an RS for services and social partners. The RS has two main modules: a recommendation module and a physical layer module. The recommendation module combines three components: recommendation-database, service/interest-matching, and service/connection time-prediction. The Recommendation database performs three main functions: (1) it collects all the information on users and service providers, including user preferences, features, and services; (2) it creates a social network page based on this information; and (3) records the historical information. The service/interest-matching (S/I-M) matches the services of the providers with user interest using different recommendation and data mining techniques. The service/connection time-prediction (S/CT-P) is used to detect a suitable time to create a D2D link, taking the quality of the wireless channel, battery capacity and the mobility of the wireless device into account. The results of (S/I-M) and (S/CT-P) are evaluated to provide recommendations by updating the social network pages. The physical layer module is responsible for managing all the resources, such as the physical link with the partner, power consumption and codeword rate, as well as providing communication recommendations to improve the performance. The authors give a detailed example to explain the role of this module.

    \subsection{Multi-Agent Algorithms}
   A multi-agent system (MAS) has two main goals: to provide rules to construct a complex system and provide techniques to coordinate each agent’s behaviour. In \cite{stone2000multiagent}, several reasons for using a multi-agent system are presented; some of these are summarized in Table \ref{table:reasons}. These considerations have led some studies to focus on the use of MAS to build recommendations systems in IoT.
  
     \begin{table*}[ht!]   
    \centering
           \caption{Reasons for using a multi-agent system.}
              \label{table:reasons}
\begin{tabular}{l p{8cm}}
  \toprule
Reason & Explanation \\ \midrule
        Speed    & Using a multi-agent system provides
several mechanisms for parallel
computation, which increases the
system's efficiency.\\  
    
    Parallelism   &  Each agent has various tasks or abilities. \\
    Scalability   &  Some systems need to be more flexible
in term of adding a new agent or new task
to the system.\\
    Robustness  & Robustness is achieved by eliminating the single failure point.\\
    Simpler programming&  It is easy for programmers to tackle and
control subtasks rather than tackle the
whole systems.\\\bottomrule
                        \end{tabular}
\end{table*}

  In \cite{forestiero2017multi}, a multi-agent algorithm is exploited to improve the speed of the recommendations, with each thing described by thing descriptors (i.e. bit vector). Things with similar features will then be brought together and managed only by neighbour’s cyber agents. Note that all cyber agents are connected by hops. The probability function is used to decide if the thing descriptor will be moved towards a linked cyber agent. The cyber agent evaluates the thing descriptor if it arrives from another linked cyber agent. The algorithm is designed to make it both intuitive and straightforward to produce a recommendable thing. When a query is established, a cyber agent matches thing descriptors and then forwards the query to a neighbour cyber agent with a maximum value of similarity. This process continues with other cyber agents until the node agent has a maximum similarity result, compared with the current agent, and the query process then finishes. After that, the query is forwarded to the asking cyber agent to produce the user’s recommendations. The algorithm was tested and evaluated, and the results showed that the recommendations were produced faster.

The authors in \cite{di2016architecture} exploited the IoT data to build a \textbf{Mobility Recommender System (MRS)} for parking, especially in urban areas, by incorporating features such as an ideal door-to-door route. The system has two main parts: the data sources, which are divided into static and dynamic data as input to the system; and the recommender system, which is responsible for dealing with these data. A MAS is exploited here to design a mobility query engine (MQE) that can distribute IoT data based on location, type, and complexity. The goal of using a MAS is to recommend parking spaces based on the consideration of different input data. For example, when the door-to-door route requires the user to use public transport, one agent who is aware of city policy may provide sets of recommendations to the route calculation planner (RCP) that take account the city’s regulations. Another agent who knows about the user’s parking preferences may also provide several recommendations to the RCP. Finally, the MRS is responsible for collecting and aggregating the queries into a ranked list of recommendations that will be considered in subsequent recommendations, where a utility RS is used to calculate user preferences. 

Twardowski et al. \cite{twardowski2015iot} built an RS that uses data from mobile devices and other IoT devices to provide personalized recommendations in real-time. The system consists of two main parts: the Big Data Server Side and Mobile Devices Side. The Big Data Server Side is responsible for the collecting and processing of mobile devices by using a multi-agent system and the Mobile Devices Side, which combines the numbers of mobile devices, such as smartphone, sensors, and beacons. The main features of this system are that it uses edge services to reduce direct communication between the mobile devices and the server, deals with the sparse data by using matrix and tensor factorization techniques and lastly, addresses the cold start problem by using the current context information and calculating recommendations in real-time. 

The authors in \cite{jimenez2019multi} proposed an RS based on the multiagent technique to optimize electricity consumption and save costs in a smart home. The system consists of three main modules with specific agents: a device module to collect data from each appliance; a crawler module, which contains the information about electricity prices, and a recommendation module that uses data from the other two modules to suggest recommendations. There is also a control agent which manages the interaction between the modules. The recommendations part is built, based on two conventional recommendation techniques, which are KB and UB. However, the system has only been evaluated in a theoretical case study using the UKDALE dataset.

\subsection{Recommendations with Graph Database Model}
    
A graphical database model is defined to represent a structural schema. Models and diagrams, such as entity relationship diagrams, are tools that designers/architects can use to test the different data structure relationships or display validity constraints graphically before implementation \cite{angles2008survey}. Graph data models are used in applications that consider the importance of the data and the relations among data at the same level. There are several benefits of using graphs as a modelling tool \cite{guting1994graphdb}: 
    \begin{itemize}

\item It allows the user to show all the information about an object in a single node, while the related information is referred to by arcs. 
\item Queries can be referred directly to a graph structure. 
\item Graph databases can provide efficient graph algorithms to investigate specific operations and efficient graph storage structures. 
\end{itemize}
The authors in \cite{palaiokrassas2017iot} exploited a Neo4j graph database to address one of the main challenges in IoT, namely, big data management. The sensory data from a smart city, which are used to produce service recommendations, are stored in a graph database. The proposed architecture has three main components: node-red, Neo4j graph database, and the recommender. Node-Red combines two parts: (1) data source flow, which is responsible for processing all data before they are transferred to the next part; (2) heating manager flow, which uses the received data to control the heating schedule for each house, as well as enabling users to interact with the system. The data stored in the Neo4j graph database define the relations among them in order to answer the user’s questions. Finally, the RS exploits not only the predefined existing data (historical data) but also the real-time data stored in Neo4j. CF is suggested as an approach to defining the similarity among the users by using similarity metrics (Euclidean and cosine). The main feature of the architecture is that each component is independent, which enables the extension and scalability of the system’s features.

Also, the author in \cite{noirie2017towards} exploited graph techniques to build an RS which provides IoT services recommendations to the user based on their own IoT devices. The system combines three main steps: IoT services modelling, IoT service catalogue (classification algorithm) and service matching algorithms (Recommendations). The IoT services model their functions by using Typed Attributed Graphs to describe all the objects and their relations. IoT services are catalogued by using an algorithm that entails the following three steps: (1) Defining physical interfaces and their location in spaces in order to build the space’s profile; (2) determining how many smart spaces share the same profile in order to distinguish the relations between the objects; and (3), a signature is defined for the list of profiles, based on the result of step 2, with the spaces that share each profile. Service matching algorithms are used to correlate between the user request and signature of the service classes. The authors discuss two different scenarios where the system can use these algorithms. One is when the user is an expert, which means he can add a new service to the catalogue, and the other is when the system can make a service recommendation based on the requests of a typical user.    

    \subsection{Recommendations with Machine learning }
     Machine Learning (ML) can be generally considered a sub-field of Artificial Intelligence (AI). It entails the use of computers to simulate human learning and can collect and use real-world knowledge to enhance performance. The field of ML has various definitions that are informed by its application across different fields of knowledge. A general and widely applicable description is, however, apposite: "A computer program is said to learn from experience (E) with respect to some class of tasks (T) and performance measure (P), if its performance at tasks in T, as measured by P, improves with experience E" \cite{mitchell1997machine}. ML algorithms can be divided into three (supervised, unsupervised, and semi-supervised), based on the nature of the data involved, or into two (transductive and inductive learning), based on the concept learning perspective \cite{el2015machine}.

ML algorithms can be utilized to optimize the ability of traditional RSs to provide accurate recommendations to the user. In \cite{sewak2016iot}, the authors designed the Optimal State based Recommender (OSR) System by exploiting some machine learning algorithms, including Distributed Kalman Filters, Distributed Mini-Batch SGD (Stochastic Gradient Descent), Distributed Alternating Least Square (ALS) based classifier, and some ML platforms. It shifted conventional recommendations, based on user/item preferences only, into accurate recommendations that deal with real-time data. Some conventional recommendation techniques cannot be directly adapted for use in the design of RSs. Accordingly, some studies have focused on improving some of these techniques or functions. For example, the authors in \cite{guo2018mobile} proposed a framework to build an e-commerce RS that exploited the multi-sources of information to produce accurate recommendations. The framework has three main components: data sources, recommendation evidence weight, and fusion decision. The data source component uses microformats to provide a unified representation and platform for information about the mobile user. 
Recommendation evidence weight uses an improved radial basis function (RBF) neural network to define the weights of recommendations. With the improved function, the weight of each piece of evidence is easy to evaluate. In the final component, the Dempster-Shafer (DS) evidence theory has also been improved to fuse information and power spectrum to provide accurate recommendations. The framework was adapted for the exploitation of multi-source information that assists women with online clothes shopping. The results showed that exploiting multi-source information had a significant impact. For example, the probability of recommending shoes increased from only 12\% with the traditional method to 85\% with the proposed method. 

In \cite{asthana2017recommendation}, the authors exploited an ML classifier to build a recommendation engine that provides personalized wearable technologies recommendations for proactive monitoring. This approach consists of three main models: (1) the classifier model, which is responsible for predicting at-risk diseases and suggesting some measurements for each person; (2) the optimization model, which is based on the measurement results and gives suitable wearable technology based on the previous measurements; (3) the Monitoring Framework, which is responsible for monitoring the readings from all the recommended devices and sending these readings to the classifier model to update its measurements.

Moreover, the K-Means algorithm \cite{amoretti2017utravel}  is adapted in the UTrave RS application, which clusters user profiles to recommend points of interest for the user. The building of the UTrave application was based on Universal Profiling and Recommendation (UPR) that combine two main steps, which are creating the user profiles and filtering. It is evaluated by two steps: (1) a simulated user to test the accuracy of the clustering user profiles and (2) real users to verify the whole system. In another experiment, Rasch et al. \cite{rasch2014unsupervised} adapted unsupervised learning to build an RS for smart homes. The system learns user patterns and conducts recommendations based on user contexts. The authors divided the system into two main phases: a training phase, with a sequence of sensors events as input; and a recommendations phase where the input is the current user context. The proposed RS is evaluated by using two publicly available datasets that are considered as a single person. Meanwhile, a decision tree \cite{yoo2018mining}  is used to build a system which provides lifecare recommendations. The system combines the two main parts: a peer-to-peer dataset and adaptive decision feedback. In the first part, the collection of the data is based on the peer-to-peer networking environment, Open API and biosensors, after which a decision tree is used to defined and classify the data. The adaptive decision feedback part plays an important role in providing flexible results of the recommendations and facilitating the real-time lifecare services. In a more comprehensive study, Valtolina et al. \cite{valtolina2014user} considered both the decision tree algorithm and social network to propose a multi-level RS. Others \cite{rizvi2018aspire} adapted the Analytic Hierarchy Process (AHP) to build an RS for car parking recommendations. Simulation parking is used to test the system, where the result outperforms other modules using the same simulation. In another study, \cite{ayata2018emotion} ML algorithms were adapted to design an RS for music recommendations that was more accurate than traditional ones. The system used sensory data from wearable devices to detect the emotions of each user and then used this data as a source to conduct accurate recommendations. Three kinds of ML are used to classify the collected data into the target emotion, which are: random forest, kNN and decision tree.

     \subsection{Recommendations with Deep Learning}

Deep Learning (DL) is generally considered to be an extension of ML. It applies two main steps: adding multiple layers, which increase the depth of the model, and transforming the data by using diverse functions representations and abstractions of multiple levels \cite{schmidhuber2015deep}. DL includes several components, such as activation function, convolutions, and pooling. It relies on different architectural paradigms (i.e. Multilayer Perceptron (MLP), Autoencoder (AE), Convolutional Neural Network (CNN), Recurrent Neural Network (RNN) and Restricted Boltzmann Machine (RBM)). The main advantage that enables DL to solve complex problems quickly and effectively is its ability to learn features. However, its limitations are a longer training time and the need for large datasets to describe the target problem. DL has been used in numerous applications that deal with continuous data, such as weather data. In \cite{yong2018iot}, the DL technique is utilized to build an intelligent system for a fitness club. The main goal of the system is to recommend suitable exercise or detect the users’ fitness actions by monitoring the user. The authors exploited a 3D CNN to identify fitness actions which can extract time and space features from continuous frames. The public KTH dataset is used to test action recognition by using the 3D CNN, where the test accuracy reached 0.8865. Inaccurate recommendations have a negative impact on the user. To address this issue, a deep neural MLP has been adapted to make system recommendations more effective for a smart museum \cite{hashemi2018exploiting}.

        \label{fig11}
    \subsection{ Recommendations with Reinforcement Learning}
    With the recent tremendous approaches to RS, reinforcement learning (RL) has received increasing attention. It meets two majority requirements for recommendations: 1) treating the interaction between the user and the agent as the main procedure for a recommendation; 2) learning the optimal policy to increase the cumulative reward without any predefined instructions. Some studies have started to adapt RL for their RS for the IoT. 

Massimo et al. \cite{massimo2018user,massimo2017learning} have proposed inverse reinforcement learning to model user behaviour as a way of improving the quality of recommendations. This technique learns about user decision behaviour by observing the user’s actions and groups observations to learn from a small number of samples. After the user behaviour is learned by using a linear function, known as a reward function, the recommendations are produced by matching the user’s predicted choices with a large group of observations of both the user and groups of similar users. The tourism domain applies the system in both outdoor and indoor environments. The initial results showed that IRL could learn user preferences, even when the datasets were small and noisy. Authors in \cite{gutowski2017framework} used a reinforcement learning algorithm (Upper Confidence Bounds algorithm (LinUCB)) to design a framework that provides context-aware recommendations in a smart city. Also, \cite{oyeleke2018situ}, the authors proposed a system, SML, to monitor the daily indoor activities of seniors with mild cognitive impairment. This aims to recommend the correct sequence of tasks for each activity so that the user can reach his goal.

   \section{Applications of RSIoT}
    \label{sec:Applications}
    
   RSIoTs are used in a variety of application domains, including smart homes, smart health, smart car parks, personal and social, and smart tourism. This section reviews the development of some of these applications.
\subsection{Smart m-health} 

\begin{figure}[!ht]
        \centering
        \includegraphics[width = 0.3\linewidth]{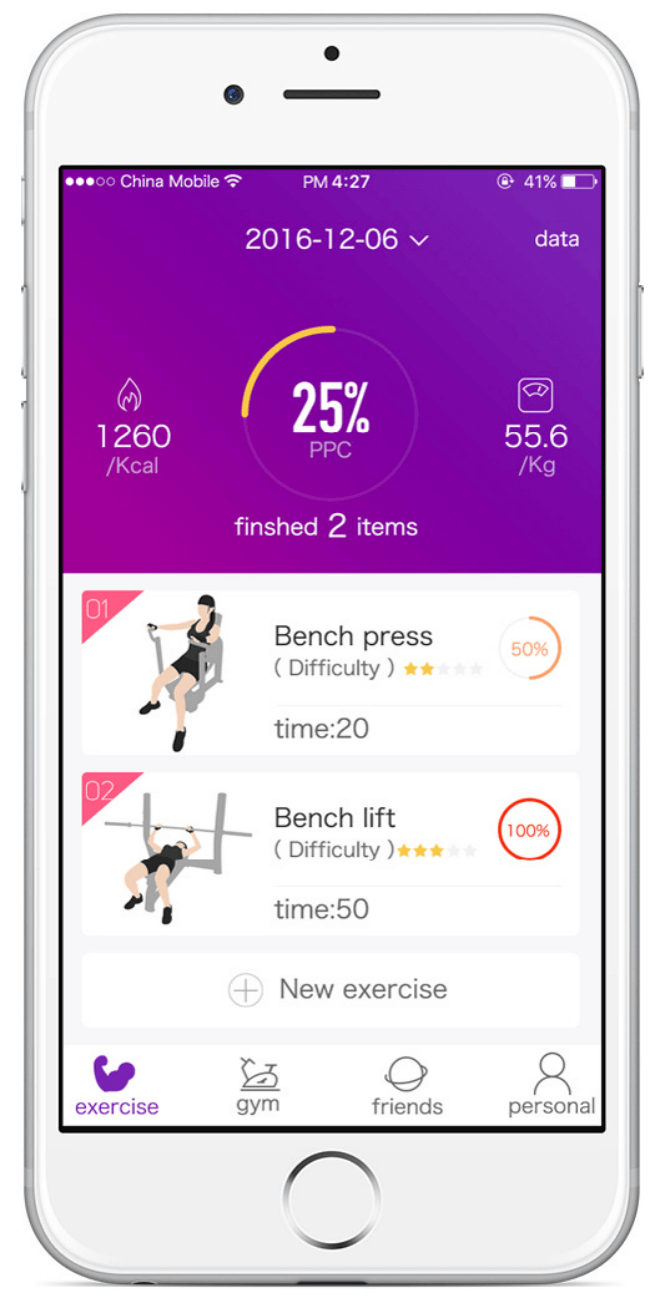}
        \caption{The homepage of our fitness application \cite{yong2018iot}.
        }
        \label{fig25}
    \end{figure}

\begin{figure}[!ht]
        \centering
        \includegraphics[width = 0.7\linewidth]{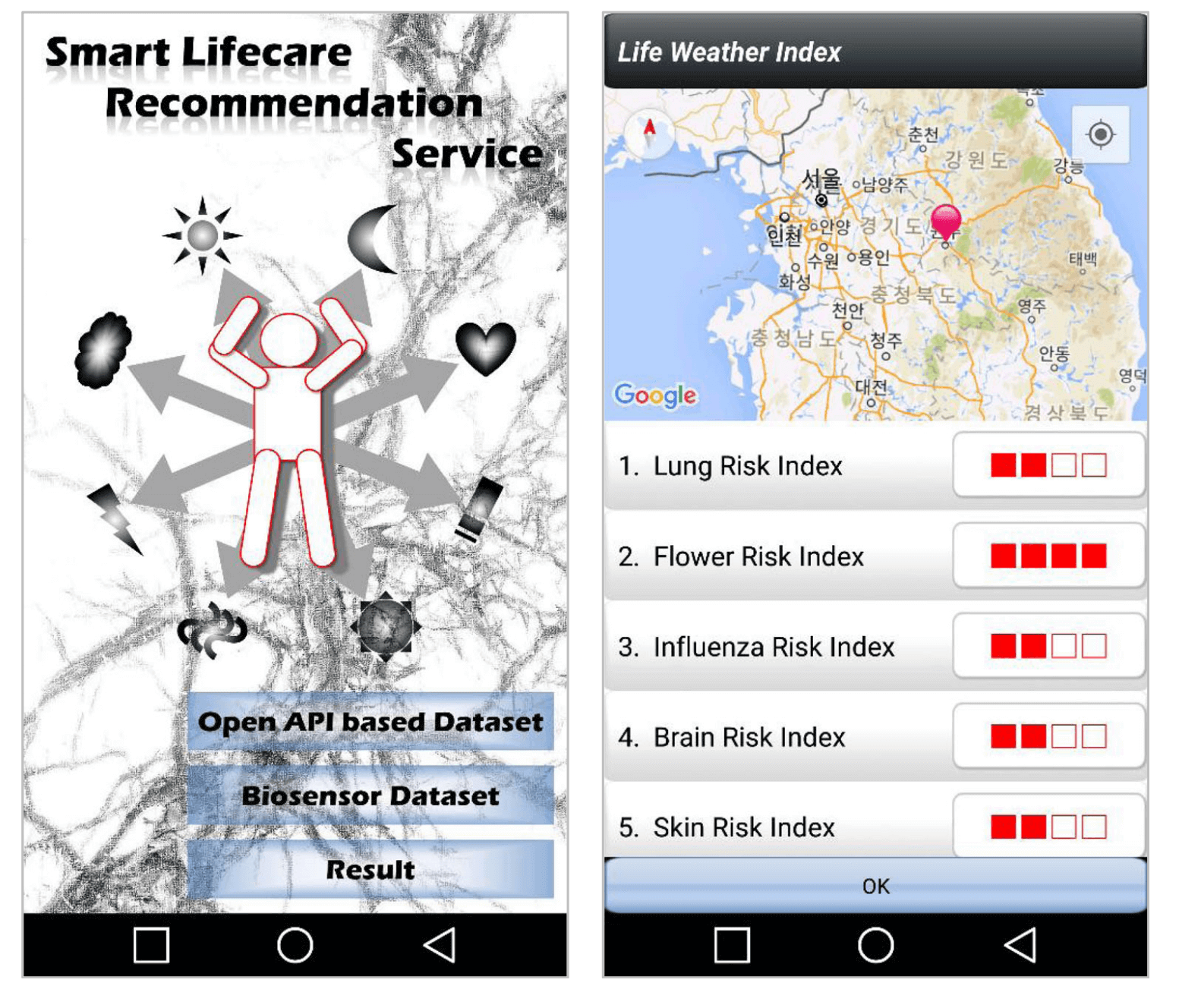}
        \caption{Initial screen and inquiry
screen of lifecare recommendation mobile service \cite{yoo2018mining}.
        }
        \label{fig26}
    \end{figure}

Over the last few years, m-health has attracted the attention of researchers who have seen it as a potential way of combining IoT with RSs to provide long-term healthcare. As a consequence, there are now numerous applications. In \cite {twardowski2015iot}, the data from mobile devices and other IoT devices surrounding them were exploited to provide personalized recommendations for dietary/fitness. The idea of this mobile application is that a higher-level context is calculated after the data is collected from smartphones and other sensors. Then, accurate recommendations are produced to match the user’s needs. For example, when the user has finished an exercise class, the system recommends a suitable restaurant that could provide a healthy meal. Yong et al. \cite{yong2018iot} designed an intelligent application to guide users in gyms (shown in figure \ref{fig25}). This application includes a part responsible for providing course reminder recommendations that are based on the users’ fitness. Another study \cite{yoo2018mining} designed a lifecare recommendation mobile service which improves the quality of life, as shown in figure \ref{fig26}.

    \subsection{Smart Car-parking}\label{sub:smart car park applications}
\begin{figure}[!ht]
        \centering
        \includegraphics[width = 0.85\linewidth]{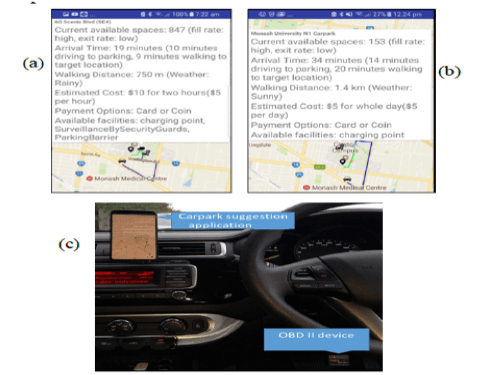}
        \caption{Smart Carpark Recommender PoC \cite{hassani2018querying}.}
        \label{fig12}
    \end{figure}

    Car park recommendation is considered to be one of the crucial services in smart cities \cite{ji2014cloud}. There is rapid growth in the design of applications for parking recommendations that exploit IoT data. In \cite{hassani2018querying}, one such application has been designed that exploits context services data owned by several providers in order to produce accurate recommendations for users. The basic idea of this application is that four context services are used as sources to produce accurate recommendations, which are described using CSDL. Also, an OBD II device is connected to the application using Bluetooth that provides some of the required sensory data, such as speed and fuel level. As can be seen in Figure \ref{fig12}, the application considers the context services before providing recommendations. For example, if there is bad weather, it considers the short walking distance in the car park recommendations (figure \ref{fig12} (a)), but if the day is sunny, the car park recommendations will consider low-cost (figure \ref{fig12} (b)). The last part of figure \ref{fig12} refers to the application running on a mobile phone and OBD II device, which is exploited as a data source in this application. 
    
\begin{figure}[!ht]
        \centering
        \includegraphics[width = 0.95\linewidth]{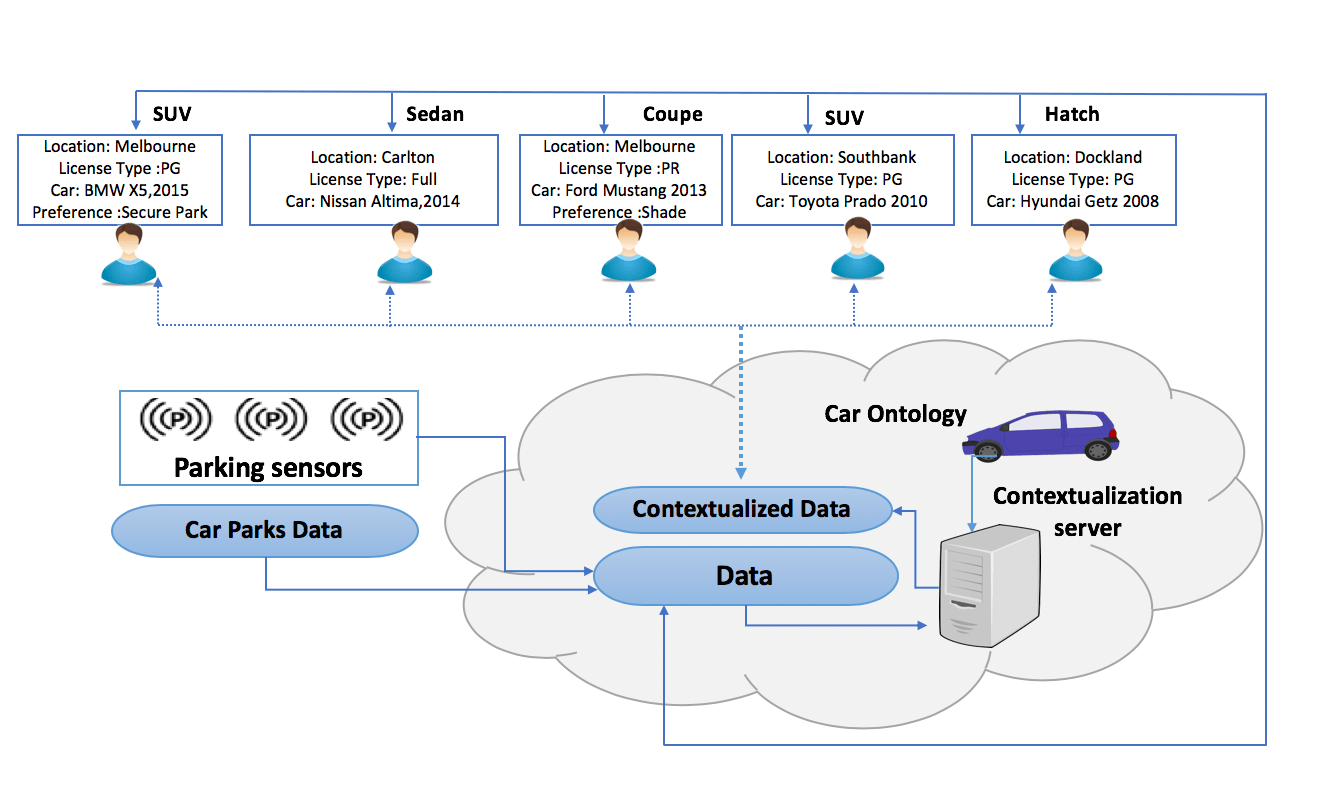}
        \caption{Contextualization Architecture \cite{yavari2016contextualised}.}
        \label{fig13}
    \end{figure}

In \cite{yavari2016contextualised}, which is another smart parking application focused on collecting contextualized data for each driver, IoT data and car park services data (see Figure \ref{fig13}), contextualization methodology is applied to the collected data to provide parking recommendations. The experiment’s results showed that the contextualization produced a three-fold reduction in query response time compared with other approaches.
Another study \cite{di2016architecture} focused on using a mobility RS where each agent was responsible for providing a list of recommendations. For example, one agent provided only direct car parking spaces based on the destination. In contrast, another agent had city policy regulations so the recommendations could be different from the previous one. The main feature of this application is that the user has several recommendations that consider his preferences from many perspectives. However, parking applications still face two main issues: using the nearest parking space, instead of parking that is recommended, and the recommended parking space being stolen. These two problems cause conflict parking. To address these problems, in  \cite{srisura2019periodical}, an application for parking conflict reduction was proposed that uses a periodical recommendation. Parking spaces that are available at different periods during the day are recommended to ensure the accuracy.

    \subsection{Smart Tourism}
    Some recent tourism literature indicates that providing tourists with a unique and differentiated service has a positive impact on marketing. When tourists explore an area, they require high-quality services that will lead to memorable experiences. However, the dramatic increase in options available to them means tourists have difficulty making decisions. E-tourism RSIoTs address this issue by providing accurate recommendations. There are several tourism RSs. For example, authors \cite{cha2016role} designed a smart tourism application by exploiting two of the main sources to provide real-time recommendations: user mobility pattern and points of interest. The application sends recommendations that are suitable for each user and based on his or her location. Also, in \cite{massimo2017learning}, a tourism application is designed which recommends points of interest to visitors. User behaviour is used as a source to reduce potentially false information that could affect the quality of the recommendations made by tourism applications. The idea of this application is that there are several media in different spaces that the visitor can watch by using the NFC tag. Each user has a specific ID, and the media are shown when the NFC tag is passed through the NFC reader. By monitoring and filtering all the interactions between the media and the visitors, the system can recommend points to visit that match the user preferences. The authors in \cite{hong2017social} proposed an RS application which provides a list of artworks recommendations that are based on calculating the social affinity between the user and user experience.

    \subsection{Personal Recommender System Applications}  
\begin{figure}[!ht]
        \centering
        \includegraphics[width = 0.85\linewidth]{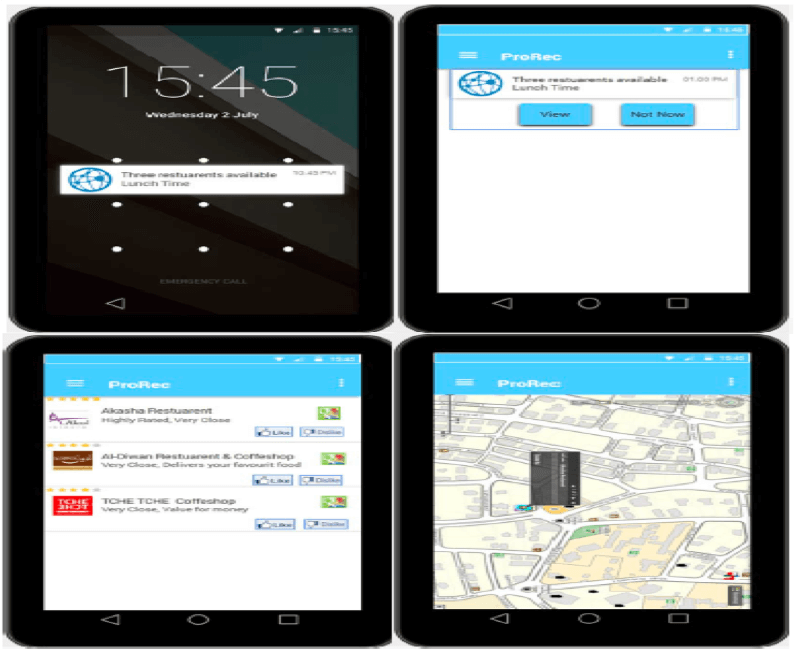}
        \caption{Screenshots for the prototype \cite{salman2017model}.}
        \label{fig14}
    \end{figure}

When a personal RS makes recommendations that resolve complex problems, the bridge between the user and objects is built. The design of this kind of system has become important, especially with IoT data. Accurate recommendations can be provided by exploiting massive amounts of IoT data and knowing when and what kinds of recommendation should be pushed. For example, an application (ProRec) \cite{salman2017model} has been designed that supports the multitype context for recommendations. Three types are used: restaurants, gas stations, and attractions. When the user downloads the application, s/he can control it by using the options setting. The recommendations will be pushed to the user, based on how close they are to the restaurants, rating, and other options (such as whether the restaurants are open at lunchtime). Figure \ref{fig14} displays different screenshots of the application. Acceptance or rejection of these recommendations will affect the results. For example, if the user does not accept the pushed recommendations, the rating of these recommendations will be reduced.

\begin{figure}[!ht]
        \centering
        \includegraphics[width = 0.85\linewidth]{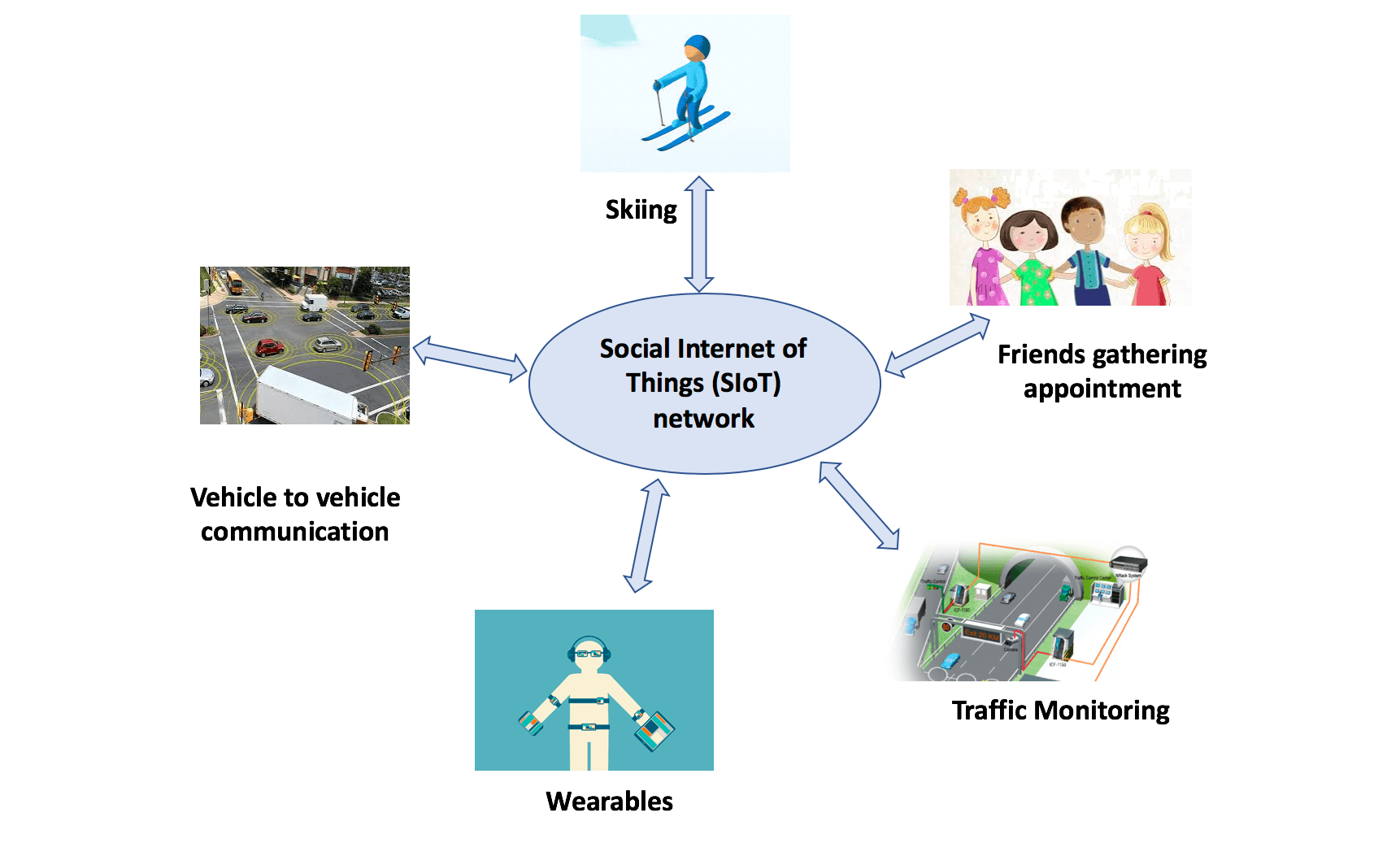}
        \caption{A sample application scenario in which different applications benefit
            from the SIoT by using other application's data \cite{saleem2016exploitation}.
        }
        \label{fig15}
    \end{figure}

   There are other applications that provide recommendations by focusing on exploiting the data of other applications. For example, the authors in \cite{saleem2016exploitation} used a sample application scenario to show how social IoT provides benefits for IoT applications by using other applications’ data. Several IoT applications are used: skiing, arranging meetings with friends, vehicle-to-vehicle communications, traffic monitoring and demonstrations of wearable (see Figure \ref{fig15}). An interoperability layer is responsible for enabling the interoperability between these applications so that the IoT RS can exploit the data to produce recommendations to the user. One of the most important features in this scenario is that IoT data can be exploited by different applications at the same time and it can investigate scalability when large amounts of things are added to the network. However, some implementation challenges are identified in the paper.
    The work in \cite{kamal2013autonomic} used the data from an e-mail survey and traffic analysis to provide personalized recommendations to the user. The idea of this application is that the user has a chance to define their choices of interest, such as emotion, location, weather and time. The system then monitors and analyses the smart devices’ traffic to predict services recommendations which match their preferences. Another application \cite{chakraverty2018iot} exploited IoT data, such as weather data, to provide several recommendations that are based on the weather. It uses sensory data to collect weather observation sequences and analyses them. Then it correlates current observations with previous ones to extract the trends that are predominant during this period and provides recommendations to the user. 
    
    \subsection{Social Recommendations}
\begin{figure}[!ht]
        \centering
        \includegraphics[width = 0.95\linewidth]{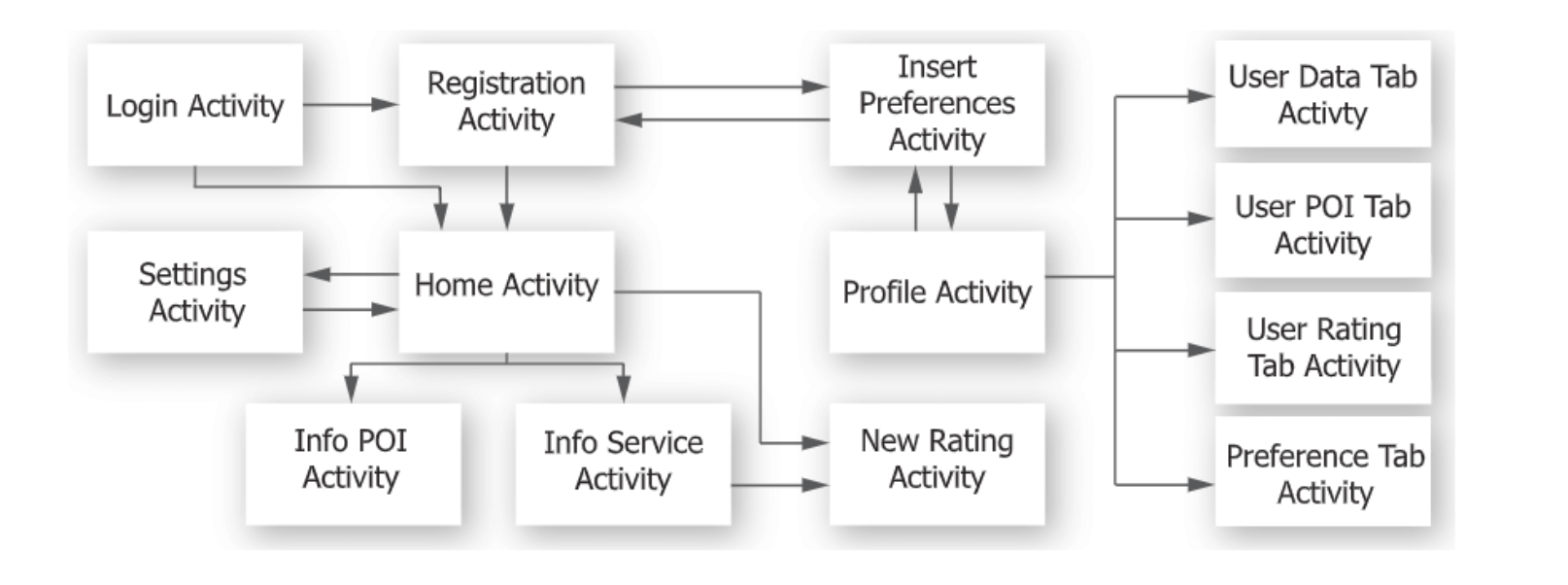}
        \caption{The structure of the UTravel mobile application \cite{amoretti2017utravel}.
        }
        \label{fig67}
    \end{figure}

  Social recommender systems not only conduct item recommendations based on social network data but also help to create relationships between users based on their preferences \cite{manca2014design}. Most applications for IoT are focused on exploiting social data to correctly define user preferences. For example, Amoretti et al. \cite{amoretti2017utravel} built a UTrave application that recommends points of interest for visitors (shown in figure \ref{fig67}). This application consists of a mobile application side and a server side. The former collects personal data by using applications installed on the client’s mobile devices; the latter leverages the collected data. The main feature of the application is that it could deal with the cold start problem – when the user sends a request recommendation for a new POI to the server but it is not available, so the server starts to use the online data sources of the user (Demographic data) to recommend the nearest POIs from the user. Yuan et al. \cite{yuan2015and}exploited some of the IoT data, such as tweet data (time, user, words, Geo), to provide accurate recommendations. 
  
    \subsection{Smart Homes} 
        \begin{figure}[!ht]
        \centering
        \includegraphics[width = 0.4\linewidth]{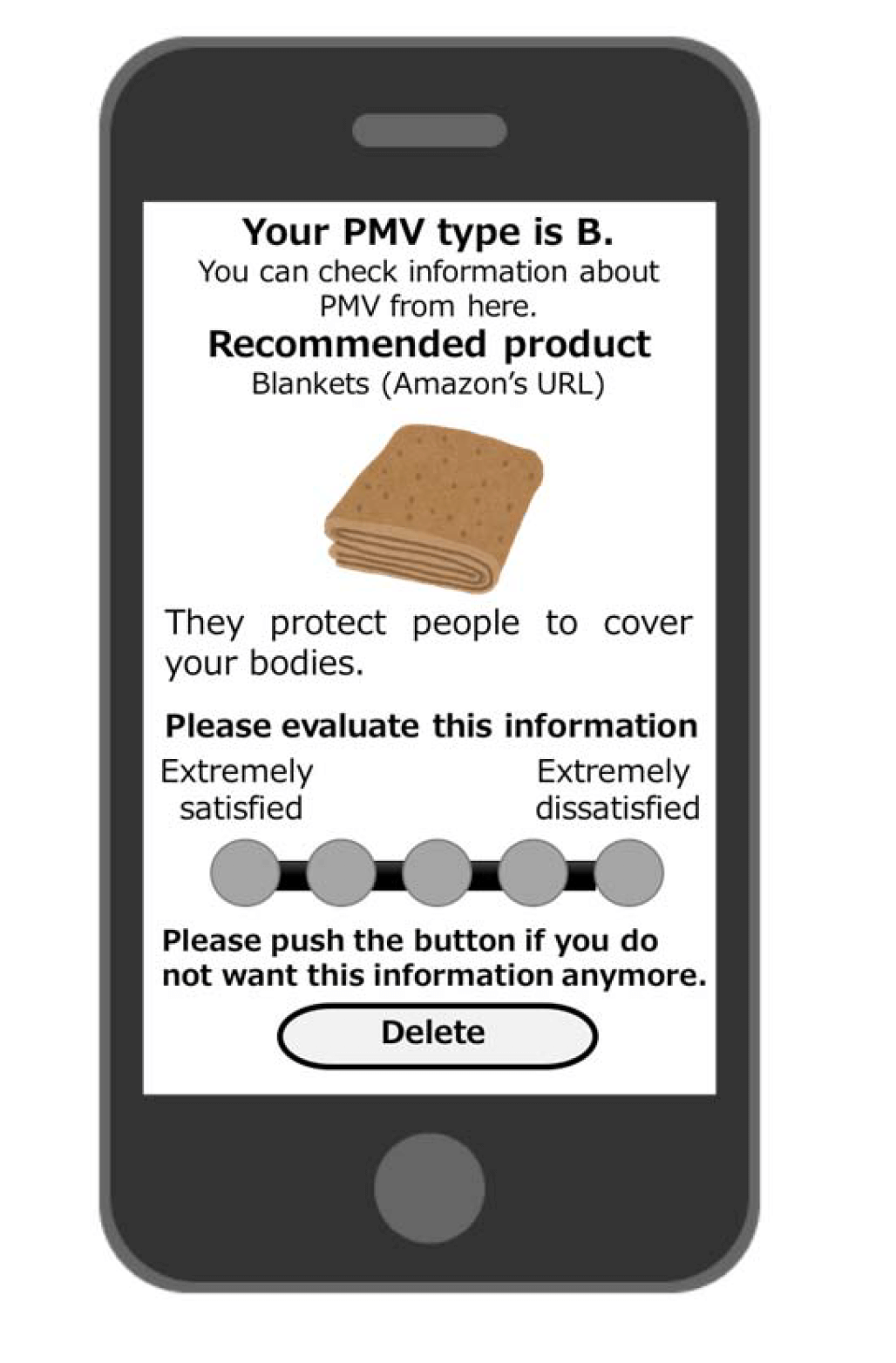}
        \caption{Screenshot of the webpage. \cite{matsui2017recommendation}.
        }
        \label{fig24}
    \end{figure}
   In the last few years, smart home applications have become increasingly popular; consequently, RS applications for smart homes have become available that help the user to choose the option that matches her/his preferences. For example, a heating schedule management application \cite{palaiokrassas2017iot} is designed that exploits open data from the city. These are gathered by using the IoT infrastructure to provide heating recommendations for a specific user. The application considers several factors while compiling the list of recommendations. The user, for instance, can choose an application that automatically turns the heating on during a specific period. For example, a resident with a flat of 60 $m^2$  in a specific location might want to apply economic heating for the next ten hours. Another application \cite{yao2014exploring} has exploited three kinds of correlations to produce accurate home recommendations: the relationships between users, the correlations among things and the interaction between users and things. Also in \cite{felfernig2017recommendation}, the authors exploited conventional recommendation approaches for AGILE projects. These help the user to choose several kinds of recommendations; for example, recommending a suitable application for an installed device, suggesting some devices that may be required for some applications, or recommending some specific protocols which match devices. A recommender system \cite{matsui2017recommendation} was designed that provides indoor comfort products recommendations for users. The system is divided into two main parts: the first is the data collecting part, achieved by using a network of sensors and extracting user preferences, while the second part involves using previous information to make product recommendations. Figure \ref{fig24} shows a screenshot of the web page of the system, which provides the product recommendation at the top of the screen and the evaluation and delete button at the bottom.

    \subsection{Smart Marketing}
    
    In recent years, online shopping RSs have been developed that help customers by providing various guidelines. Most of these systems use conventional recommender approaches to providing recommendations for large websites, such as Amazon and eBay. However, with IoT, the recommendations could be made more attractive by exploiting IoT data as sources. A shopping centre application \cite{saghiri2018framework} has been designed that recommends healthy products to the customer by exploiting sensory data. The customer can also pay automatically using a digital wallet. An RS \cite{tu2016context} is proposed which makes DS more interactive. The system exploits the interaction between the user and the DS to make the recommendations.

\section{Analysis and Findings}
 \label{sec:Analysis}
\begin{figure*}[ht]
\begin{minipage}[b]{0.45\linewidth}
\centering
\includegraphics[width=\textwidth]{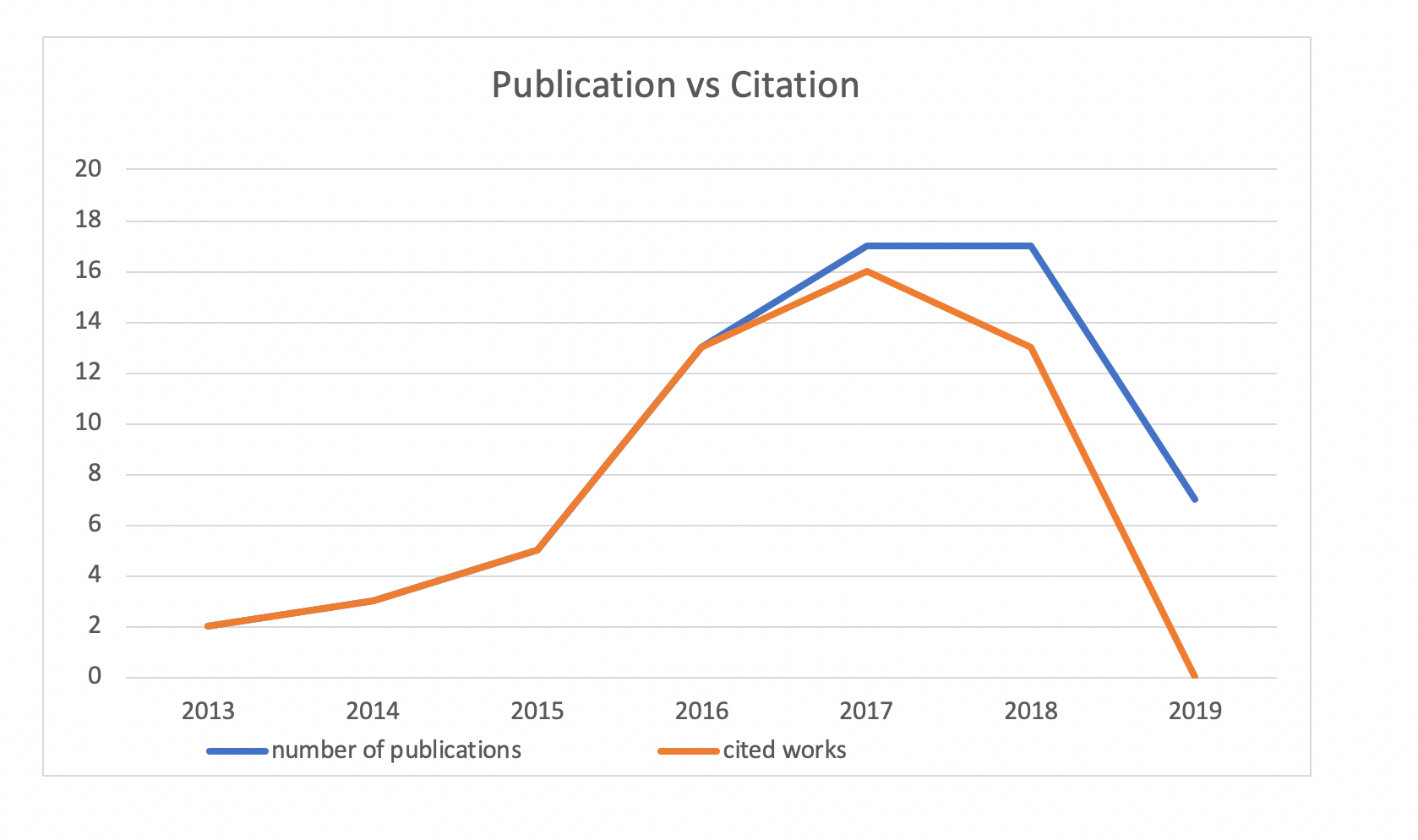}
\caption{Number of publications, in-field cited works, and in-field citations.}
\label{fig:figure1}
\end{minipage}
\hspace{0.5cm}
\begin{minipage}[b]{0.45\linewidth}
\centering
\includegraphics[width=\textwidth]{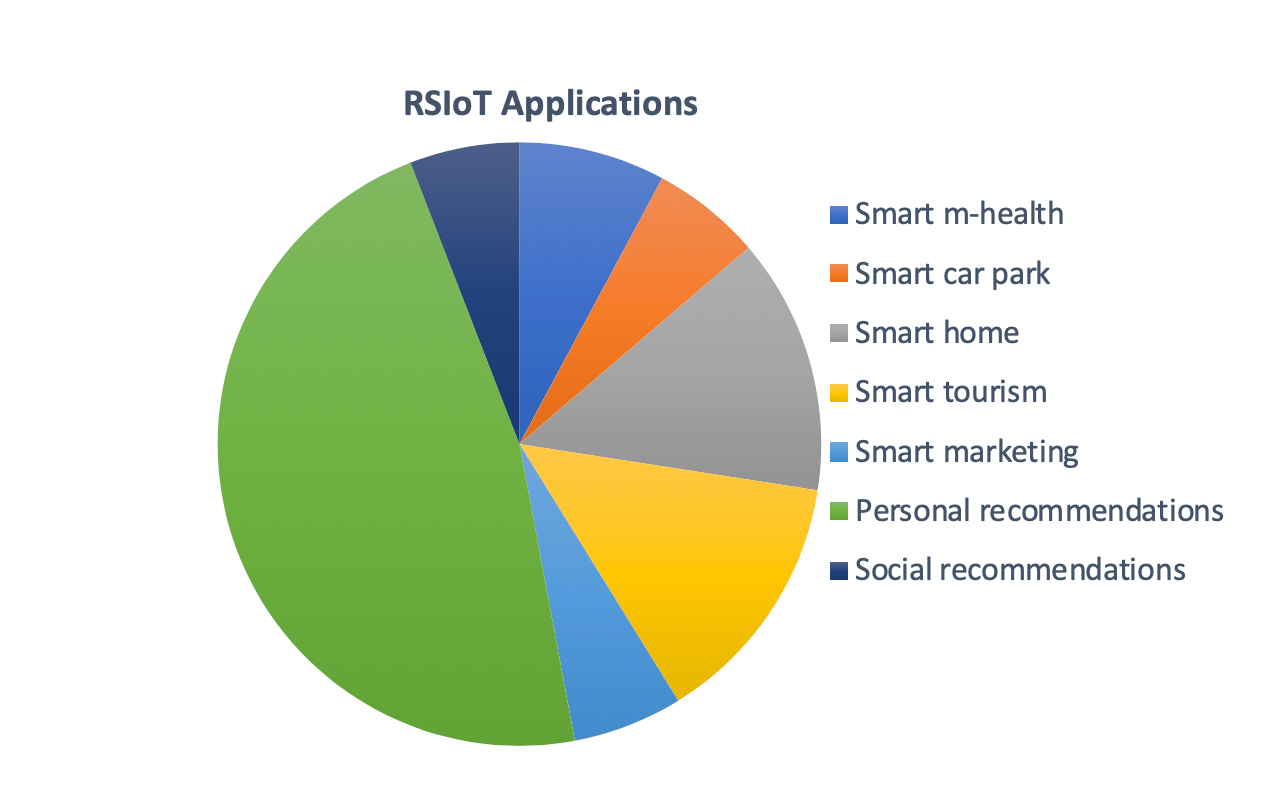}
\caption{Proposition of RSIoT applications.}
\label{fig:figure2}
\end{minipage}
\end{figure*}

Figure \ref{fig:figure1} shows the number of published works in two fields: cited works and citations for each year, from 2013 to 2019. The number of works that are related to RSIoT has increased steadily each year since 2013, with a surge of interest in 2016. This surge represents a gradual shift in focus towards combining the internet of things with an RS. It also reflects the sharp perception of designing RSs based on real-time recommendations. However, the number of in-field cited works have not kept up with the number of publications; particularly in the last two years. The main reason is that the majority of works might still not be detected by researchers, compared with those in other years, such as 2016 and 2017. While citation count is not a perfect metric to evaluate the impact of research works, it does, however, show the extent to which a researcher’s work is being acknowledged.

The following part reviews RSIoTs and their applications. Table \ref{tab:summerized} shows the main recommendations techniques and applications. The main trends in RSs for the IoT have been reviewed through studies published from 2013-2019. Some of the studies exploited conventional recommendation approaches, such as CF, CB, UB, and KB, in RSIoTs, but improvements in these approaches have addressed some of their limitations, which has had a positive impact on the development of other RSs. However, as we mentioned in Alice’s scenario, RSIoT should consider more than the interaction between users and items. For example, our RS will not recommend a cup of coffee to Alice just because she has entered her kitchen. It needs to collect more information before making any recommendation, such as time. Accordingly, using traditional approaches to make the recommendation could be inefficient for RSIoT. The RSIoT, based on the techniques that we discussed above, such as context aware, Social IoT, multi-agent and graph techniques, started to shift from recommendations that depend only on the interaction between the user and item as a resource to recommendations with more resources. For example, context technique considers any information about the user and item, such as identity, location, state of people, etc., which provide rich input that helps an RS to make accurate recommendations.

However, the previous techniques were not able to learn the human pattern. In Alice’s scenario, our RS should consider Alice’s daily pattern; so preparing a cup of coffee in the morning would not be recommended if Alice changed her habit and it will consider a new activity instead. The last three techniques, which are machine learning, deep learning and reinforcement learning, provide promising directions for RSIoT. We discussed in section \ref{intro} that exploiting the knowledge of human pattern plays an important role to conduct accurate recommendations. In one of our previous studies \cite{altulyan2019reminder}, we proposed a reminder care system which focused on detecting the complex activities for the user and then providing reminder recommendations for patients with Alzheimer’s disease. The framework is divided into three main stages: (1) Correlation analysis of devices, (2) Rule-based orchestration, and (3) Activity-triggered Recommendation (see Figure\ref{fig00}). We adapted one of the deep learning algorithms, DeepConvLSTM, which is a state-of-the-art deep neural network that combines convolutional, recurrent and softmax layers for the first stage. ML algorithms could also be adapted for this task. However, the main challenge is how the RS can learn the human activity pattern in order to push the recommendations that match the user preferences. Each user has a different pattern that is compared with and this pattern could be changed after a certain period; the user prefers to drink a cup of coffee in the morning during the weekday but prefers to sleep in at the weekend. The deep reinforcement learning algorithm can address this issue because it has the ability to capture the user’s temporal intentions and make recommendations in a timely manner. Regarding the applications, we discussed six major RSIoT applications that mostly fell in the smart home and personal domains, as shown in figure \ref{fig:figure2}.

    \section{A Unified Recommender System Framework for the IoT and Future research Directions}
     \label{sec:Unified}
    
Defining a unified RSIoT would be of great benefit to researchers and professionals. Figure \ref{fig16} presents such a framework that is based on state-of-the-art knowledge of RSIoT. The framework focuses on the following four main stages:

\begin{enumerate}
      \item\textbf{Data acquisition platform}
         
 Preparing an efficient platform for sources that are used to feed the RSIoT is crucial. There are three basic steps required to create an RSIoT platform. The first step is to turn physical things into smart objects. We can define smart objects as objects that can provide information and data about themselves or other related objects. They are also capable of communicating this information \cite{haller2002need}. There are two mainstream ways to turn physical objects into smart things: tagging the object with RFID tags and embedding the sensors. However, the most distinctive issue is how to deal with the resource constraint of physical things. To address this problem, it is feasible to embed each physical thing to a server, so that they each have communication, storage, computation, and data processing abilities and can provide services (e.g. smartphones, sensor nodes, and RFID readers). RFID and Wireless Sensor Networks are two main enabling technologies for bridging physical things to the virtual world. In the next step, every smart device (thing) should be connected to the Internet. Things can be connected, wired or wirelessly. In RSIoT, a wireless connection will mainly be used. There are many ways to connect a smart device (thing). Based on the existing infrastructure, these are RFID, ZigBee, WPAN, WSN, DSL, UMTS, GPRS, Wi-Fi, WiMax, LAN, WAN, 3G, etc. The last step is to employ specific network protocols that are used in sensor networks and RFID networks to make them connect and communicate via the Internet. A number of IoT protocols have been proposed which consider sensor restrictions, such as limited energy and computation capability. For example, Message Queue Telemetry Transport (MQTT) \cite{website2014} is used for M2M and IoT applications that need a lightweight subscribe and publish message. The Constrained Application Protocol (CoAP) \cite{shelby2012constrained} works as a web transfer protocol for constrained nodes and M2M applications. Other IoT protocols for outing/forwarding have also been proposed that are designed for peer-to-peer communications or client-server communications, such as IPv6 over Low Power Wireless Personal Area Network (6LowPAN) \cite{schumacher2007ipv6} and WebSocket \cite{fette2011websocket}. Sensor networks and RFID have their own communication stack; 802.15.4 is widely used for sensor networks, while RFID usually supports the 802.11 standards to the wireless network on the Internet.

\item\textbf{ Data processing for rich information }

  The data processing phase is required for rich information, as it ensures that the data can be used effectively for further processes. It has a critical impact on the accuracy of results and is therefore crucial for the success of the whole system. After the data is collected, as we mentioned previously, it may contain noise, missing values, and redundant features. Consequently, a number of sensor data processing methods can be adopted, which are classified into three main parts: reprocessing, segmentation, and dimensionality reduction. The first part includes operations to clean the data, dealing with missing readings and transforming the data. Usually, the sensor’s readings are converted from one measurement to another, such as from voltage to temperature, and this process can affect the data \cite{aggarwal2013internet}. In order to tackle this problem, the sensor is recalibrated \cite{bychkovskiy2003collaborative} or adapted via data-driven modelling \cite{deshpande2004model}. In addition, the data from RFID is nosier than the data from a sensor, because of errors related to reader tag communications and redundant data. Several techniques can be used to clean RFID data \cite{aggarwal2013managing} or deal with losing RFID data \cite{gupta2004developing}. In the segmentation part, the raw data from sensors continuously flow and therefore needs to be divided into smaller parts. The selection of the proper segmentation approaches has had a huge effect on the last part of the data processing. The authors in \cite{ni2015elderly} categorized the segmentation approaches into three main types: temporal-based segmentation, activity-based segmentation and sensor event-based segmentation. Dimensionality reduction plays the main role in extracting and selecting features from the raw data by using different approaches. In \cite{figo2010preprocessing}, the authors proposed pre-processing mechanisms that could handle any errors of raw data, specifically, accelerometer sensors. Sensor signal processing techniques are classified into three main domains: the time domain, which contains mathematical and statistical metrics to extract the basic signal information from sensory data; the frequency domain, which is responsible for capturing the repeated nature of the raw sensor data; and the discrete representation domain, which transforms the signal sensor into a string of discrete symbols.

\item\textbf{Event generator and Rule composer}

In a system that recommends things to users, we intuitively need to first understand what kinds of things the user prefers. There are two main tasks that should be implemented at this stage: (1) generating the events based on previous information, and (2) defining a suitable rule for each event. In the first task, RSIoT should be able to extract an event from the extracted features at an earlier stage. The event depends on the kinds of applications and the extracted features. For example, in Alice’s scenario, the system can detect that she prepares a cup of coffee in the kitchen at midnight. Three data sources are exploited to extract this complex activity event: activity recognition (standing), localization (the kitchen), and object usage detection (the cup of coffee). Several probability-based algorithms have been used to implement this task. The HMM can be adapted for activity recognition \cite{kim2010human}. It can generate a hidden state, based on the observable data, and learn reliable model parameters from the history of the model output. However, it still has some limitations, such as in representing the number of interactive activities, capturing the long-range of the observations, and the failure in recognizing the available observations for a consistent activity \cite{gu2010pattern}. Another model that could also be used is the conditional random field (CRF) \cite{sutton2012introduction}. This is considered to be more flexible than the HMM because it focuses on extracting the conditional probability instead of a joint probability distribution, such as HMM. ML models have been used for activity recognition such as Naive Bayes classifiers \cite{logan2007long,tapia2004activity} and decision trees \cite{maurer2006activity}. Ontological models also help to recognize complex activities \cite {yamada2007applying,altulyan2019reminder}. In Alice’s scenario, an ontological model could be built to represent human household activities and environmental domain concepts and objects. Another task for this stage is a rule composer. This contains a set of rules, extended from the previous task, and combined constraints. It is responsible for defining suitable recommendations based on each event. For example, Alice will not receive any recommendations about preparing coffee or certain kinds of food when she is in the kitchen at midnight. Also, the rules have a flexible nature which can be updated, based on new events that are generated by the system. A number of efforts have been made to tackle rule modelling \cite{debattista2012ontology}, build rule engines \cite{li2010ontology}, and design techniques that enable rules for automatic learning \cite{bai2008ontocbr}. 

In our previous work, we started \cite{altulyan2019reminder} to cover this stage of the framework. We proposed a reminder care system which focused on detecting complex activities for the user and then providing reminder recommendations for patients with Alzheimer’s disease. The framework is divided into three main stages: (1) Correlation analysis of devices, (2) Rule-based orchestration, and (3) Activity-triggered Recommendation (see Figure \ref{fig00}). We adapted one of the deep learning algorithms, DeepConvLSTM, which is a state-of-the-art deep neural network that combines convolutional, recurrent, and softmax layers for the first stage. Experiments showed that the behaviour recognition part of our system is effective and hence the recommendation engine is practicable.

 \begin{figure}[!ht]
        \centering
        \includegraphics[width = 0.6\linewidth]{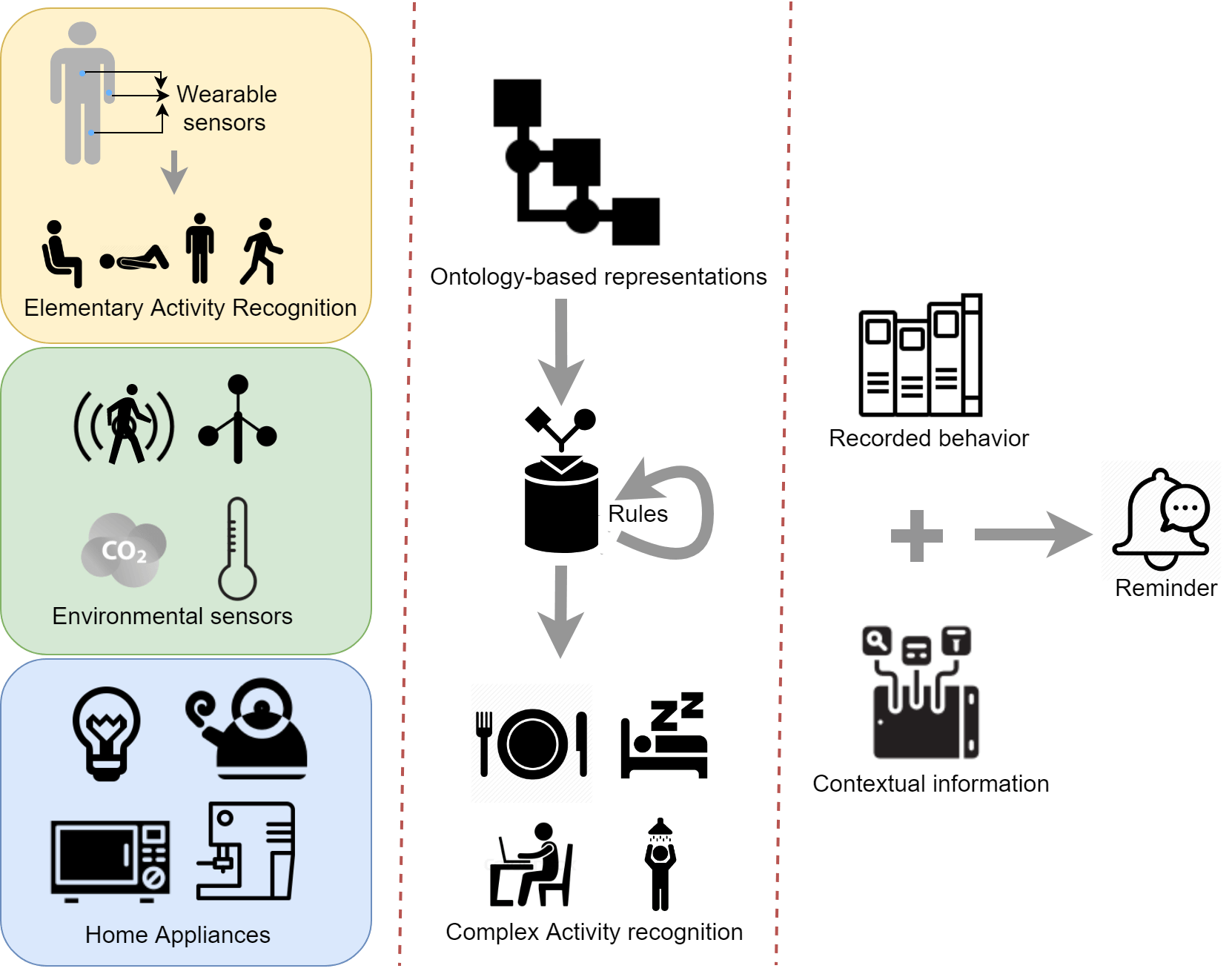}
        \caption{Reminder care system framework.}
        \label{fig00}
    \end{figure}

\item\textbf {Accurate recommendations production} 

This is considered to be the core stage of the framework because RSIoTs are less constrained and more context dependent than conventional recommendations that are built based on web data. Consequently, the techniques used to design RSIoT are more complex than those used in traditional RSs for things like books or movies. There are several reasons for this complexity, most of which are related to the limitations of IoT identified in this article. During this study, we observed that conventional recommendation approaches, such as CF, CB, UB, and KB, are adapted for RSIoTs. However. improvements in these approaches have addressed some of their limitations, which has had a positive impact on the development of other RSs. However, RSIoT should consider more than just the interaction between the users and items. For example, in Alice’s scenario, the RS will not recommend a cup of coffee to Alice just because she has entered her kitchen. It needs to collect more information before conducting any recommendation such as time.
Accordingly, using traditional approaches to conduct the recommendation might not be as efficient for RSIoT. The RSIoT, which is based on the techniques discussed above, such as context aware, Social IoT, multi-agent and graph techniques, started to shift the recommendations that depend only on the interaction between the user and item as a resource into recommendations with more resources. For example, context technique considers any information about the user and item, such as identity, location, the state of people, etc., which provide a rich input for an RS to conduct accurate recommendations. However, the previous techniques were not able to learn the human pattern. Let’s consider Alice’s scenario as an example. The RS should not conduct the recommendations only based on detecting her current status. It should consider her daily pattern as well; therefore, preparing a cup of coffee in the morning will not be recommended if Alice changes her habit, and it will consider a new activity instead. The last three techniques which are machine learning, deep learning and reinforcement learning provide promising directions for RSIoT. We discussed in section \ref{intro} that exploiting knowledge of the human pattern plays an important role in conducting accurate recommendations. Deep reinforcement learning algorithms can address this issue when it can capture the user’s temporal intentions and conduct recommendations in a timely manner. Most of the existing works focus on exploiting RL to learn the optimal recommendation strategy, which is based on the interaction between the user and agent, or to increase the cumulative reward for each scenario. However, adapting reinforcement learning for RSIoT faces major challenges in term of dealing with the human pattern in different scenarios. For example, in smart homes, RSIoT will face several scenarios during the day, as conducting recommendations during the morning period should be different from other periods, such as noon or the evening. Accordingly, this issue could be handled by combining reinforcement learning with the multi-agent approach, which would capture the sequential dependency of the human pattern in different scenarios. Also, agents will have the same memory of the history of the human pattern and work collaboratively to improve the performance of the RS.

\begin{figure*}[!ht]
            \centering
            \includegraphics[width = 0.99\linewidth]{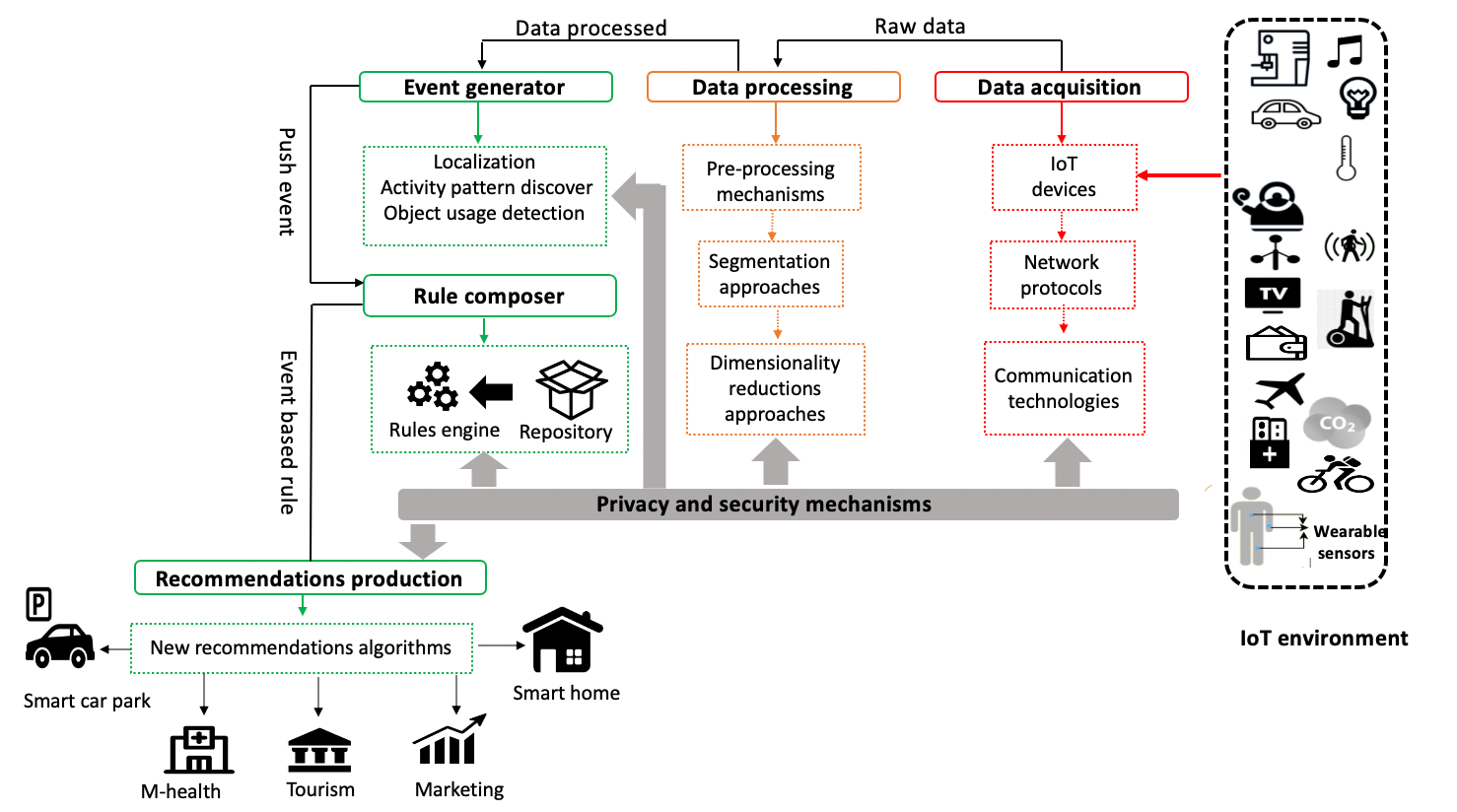}
            \caption{Proposed framework for RSIoT.}
            \label{fig16}
        \end{figure*}

         \item \textbf{Privacy and security in RSIoT }
        
The RSIoT environment deals with a massive amount of data, especially about the user in order to improve the quality of recommendations. However, collecting these kinds of data, such as health data, locations data or even daily activities, creates a number of privacy concerns for users. In addition, devices could be targeted by malicious attacks that affect the accuracy of recommendations. Consequently, several requirements and parameters need to be reckoned, which are described below. Privacy and integrity mechanisms should be adopted because IoT devices and their data are susceptible to malicious attacks that can affect or change the data. Authentication techniques, mainly for RSIoT, are also required due to the heterogeneous nature of IoT devices, and there is a need to ensure that the access system is only available to authorized devices. 
The quality-of-service (QoS) should be investigated by avoiding different attacks, such as sinkhole attacks and jamming adversaries. The efficiency of energy consumption is another requirement that needs to be considered, particularly as IoT devices have some constraints with their storage, energy and computation capability; thus, any attack on the system might increase these issues. RSIoT architecture also encompasses a wide variety of devices, starting from small devices (sensors) to large ones (servers); thus, there is a need to address security issues at different levels \cite{khan2018iot}. Recently, there has been a tremendous effort to cope with privacy and security issues in IoT. The authors in \cite{oleshchuk2009internet} explained the role of secure multi-party computations in preserving privacy for IoT users. Also \cite{zhang2014iot}, the authors explained IoT’s main security issues and highlight some future work solutions that could tackle these issues. With regard to the security issue, the studies can be classified into two parts: one targets the system’s specific security issue \cite{granjal2015security,gormucs2018security}  and the other focuses on providing security for the whole system \cite{mitchell2014survey,yi2015security}. However, addressing privacy and security issues for RSIoTs has received little attention. 
The authors in \cite{asiri2016iot} proposed an RS that investigates trust and reputation among IoT nodes by using PNN. It can provide different levels of security according to the sensitivity of the data. However, it only focuses on providing a mechanism to carefully select IoT resources regarding their reliability. In our previous study \cite{altulyan2019unified}, we proposed a holistic framework to ensure the integrity of data in smart cities, which covers the entire data lifecycle, using blockchain and other techniques. Blockchain (BC) technology can provide an efficient solution to the problem of protecting IoT recommendations. It provides a decentralized peer-to-peer network that applies stored encrypted and secure computations to the raw data. Moreover, it provides the user with authorization to access her raw data or allow another party to provide recommendations without displaying her privacy data.

        \end{enumerate}


\section{Conclusion }
 \label{sec:Conclusion}

 RSIoTs have become a crucial tool that can support users in various decision-making activities. In this article, we have provided an extensive review of the most notable works to date on RSIoTs. We have proposed a classification framework of three categories for the organizing and dividing of existing publications and then highlighted each category in more detail. We have also discussed the limitations of using IoT for recommendation tasks, as well as the most important requirements for the RSIoT. We have provided a detailed account of the techniques of each system and its applications. In addition, we have detailed some of the most common recommendations that are provided by building an RSIoT. Both IoT and RSs have been ongoing hot research topics in recent years. While existing studies have established a solid foundation for research into RSIoT, our analysis has generated several promising directions for future research:
\begin{itemize}
 \item Multi-Source Contextual Correlations: For a system to recommend things of interest to the user, we first need to understand what the user needs or what their preferences are. Multi-source contextual correlations, which include interactions between the user and things, spatial-temporal contexts, and contemporary activities, should be exploited as sources in constructing an RSIoT. 
\item RSIoTs have a highly dynamic nature compared with traditional web objects. Unlike web search engines, which are based on assumptions that most web content changes slowly, the main feature of a recommendation engine for IoT should be sufficient to discover structured and rapidly changing content. To this end, the next step would be to develop a dynamic tracking algorithm for monitoring a thing’s status, which would be a promising direction. 
\item  Trust-aware IoT Service Recommendations are required before designing a new application based on IoT, as we need to be careful about which IoT resources are reliable enough to be selected. The main issue is that traditional techniques cannot detect the reliability of IoT resources, and little attention has been paid to the trust issue of IoT resources and applications. 
\end{itemize}

Finally, we hope this survey can provide readers with a comprehensive understanding of the main aspects of this field by highlighting and explaining the most notable advancements.

\clearpage

\centering

\begin{table*}[!ht]

  \centering
\caption{Categorization by techniques type,applications domains and recommendations.}
\label{tab:summerized}



\resizebox{18cm}{!}{
\begin{tabular}{|>{\large}c|>{\large}c|>{\large}c|>{\large}c|>{\large}c|>{\large}c|}
\hline
\textbf{\large Perspective}                                                                                             & \textbf{\large Reference}                                                           & \textbf{\large Technique}                                                         & \textbf{\large Application}               & \textbf{\large Thing of interest} & \textbf{\large Key contribution}                                                                                                                                                                                                                 \\ \hline
\multirow{17}{*}{\textbf{\begin{tabular}[c]{@{}c@{}}Conventional \\ recommendations\\  techniques\end{tabular}}} & Yao et al.\cite{yao2014exploring}                                                         & \begin{tabular}[c]{@{}c@{}}CF,graph,\\ random walk\end{tabular}            & smart home                         & device                     & \begin{tabular}[c]{@{}c@{}}designs a unified CF  approach\\  based on matrix factorization \\ which exploits \\ three kinds of correlations\end{tabular}                                                                                  \\ \cline{2-6} 
                                                                                                                 & Asiri et al.\cite{asiri2016iot}                                                         & CF,PNN                                                                     & unknown                            & device                     & \begin{tabular}[c]{@{}c@{}}adapts PNN and CF\\  to propose a recommender system\\  based  trust and reputation model\end{tabular}                                                                                                         \\ \cline{2-6} 
                                                                                                                 & \begin{tabular}[c]{@{}c@{}}Chakraverty and \\ Mithal \cite{chakraverty2018iot}\end{tabular}    & CF,HMM                                                                     & personal                           & service                    & \begin{tabular}[c]{@{}c@{}}addressing two main problems:\\ model weather\\  conditions in short term\\  using HMM and\\  predict users preferences\\  using CF\end{tabular}                                                               \\ \cline{2-6} 
                                                                                                                 & Sawant et al.\cite{sawant2017representation}
                                                                                                                                                                      & CF,CB                                                                      & personal                           & service                    & \begin{tabular}[c]{@{}c@{}}combines cyber-physical systems \\ and IoT to conduct\\  recommendations based on \\ users preferences .\end{tabular}                                                                                          \\ \cline{2-6} 
                                                                                                                 & Lee and Ko \cite{lee2016service}                                                         & user-based CF                                                              & personal                           & service                    & \begin{tabular}[c]{@{}c@{}}propose user based-CF \\ to provide group recommendations  \\ by considering\\  the member organization.\end{tabular}                                                                                          \\ \cline{2-6} 
                                                                                                                 & Mashal et al.\cite{mashal2016performance}                                                        & \begin{tabular}[c]{@{}c@{}}UBCF,OBCF\\ ,bipartite graphs\end{tabular}      & personal                           & service                    & \begin{tabular}[c]{@{}c@{}}uses bipartite graphs and two algorithms\\  based on  CF to address the problem of \\ recommending IoT third party services\end{tabular}                                                                       \\ \cline{2-6} 
                                                                                                                 & Yao et al.     \cite{yao2015service}                                                          & CF                                                                         & personal                           & software                   & \begin{tabular}[c]{@{}c@{}}adapts matrix\\  factorization framework \\ to build recommender system\\  for mashup recommendations\end{tabular}                                                                                             \\ \cline{2-6} 
                                                                                                                 & Nizamkari \cite{nizamkari2017graph}                                                          & \begin{tabular}[c]{@{}c@{}}graph-based trust\\ ,CF(optimized)\end{tabular} & personal                           & device                     & \begin{tabular}[c]{@{}c@{}}Addressing the common\\  problems for traditional CF \\ by using graph-based trust\end{tabular}                                                                                                                \\ \cline{2-6} 
                                                                                                                 & Salis et al.\cite{salis2018anatomy}                                                       & ML,CF                                                                      & personal                           & service                    & \begin{tabular}[c]{@{}c@{}}adapt ML and CF to provide\\ real time recommendations\end{tabular}                                                                                                                                            \\ \cline{2-6} 
                                                                                                                 & Jabeen et al. \cite{jabeen2019iot}                                                        & \begin{tabular}[c]{@{}c@{}}Advice-based CF,\\ ML\end{tabular}              & personal                           & service                    & \begin{tabular}[c]{@{}c@{}}uses ML classification algorithms to\\  detect cardiovascular disease \\ and proposes advice-based \\ CF approach to  \\ conduct a suitable recommendations\\  based on the classification result\end{tabular} \\ \cline{2-6} 
                                                                                                                 & Li et al.\cite{li2019personalization}                                                            & CF                                                                         & personal                           & service                    & \begin{tabular}[c]{@{}c@{}}enhances the the basic matrix\\ factorization model and then \\ build the trust relevancy\\  which addresses the  data sparse  problem\end{tabular}                                                            \\ \cline{2-6} 
                                                                                                                 & \begin{tabular}[c]{@{}c@{}}Margaris and Vassilakis\\ \cite{margaris2017exploiting}\end{tabular}   & CF,QoS                                                                     & smart tourism                      & service                    & \begin{tabular}[c]{@{}c@{}}improves the quality of the recommendations \\ by combining two \\ algorithms :\\ CF and QoS\end{tabular}                                                                                                      \\ \cline{2-6} 
                                                                                                                 & Yang et al.\cite{yang2019exploring}                                                          & CF                                                                         & personal                           & service                    & \begin{tabular}[c]{@{}c@{}}proposes a location-aware\\  POI recommendation system\end{tabular}                                                                                                                                            \\ \cline{2-6} 
                                                                                                                 & Rossi et al. \cite{rossi2016towards}                                                        & CF                                                                         & \multicolumn{1}{l|}{smart tourism} & service                    & \begin{tabular}[c]{@{}c@{}}Employs CF approaches to\\  provide artworks  \\ for  both individual and\\  group visitors\end{tabular}                                                                                                       \\ \cline{2-6} 
                                                                                                                 & Erdeniz et al.\cite{erdeniz2018recommender}                                                       & CB                                                                         & personal                           & service                    & \begin{tabular}[c]{@{}c@{}}proposes a new approaches \\ to build recommender system\\ in m-health\end{tabular}                                                                                                                            \\ \cline{2-6} 
                                                                                                                 & \begin{tabular}[c]{@{}c@{}}Koubai and Bouyakoub\\ \cite{koubai2019myrestaurant} \end{tabular}      & CB                                                                         & personal                           & service                    & \begin{tabular}[c]{@{}c@{}}proposes application which facilitates \\ the work  of employees in \\ a restaurant and \\ improves users experience\end{tabular}                                                                              \\ \cline{2-6} 
                                                                                                                 & \begin{tabular}[c]{@{}c@{}}Srisura and Avatchanakorn\\ \cite{srisura2019periodical}\end{tabular} & \begin{tabular}[c]{@{}c@{}}CB,\\ Constraint-based\end{tabular}             & \multicolumn{1}{l|}{smart car park} & service                    & \begin{tabular}[c]{@{}c@{}}Proposes a smart parking recommender \\ system which address two main\\  problems that face parking \\ applications by providing\\  a periodical recommendation\end{tabular}                                 
\\ \hline
\end{tabular}}
 \end{table*}

\clearpage

\begin{table*}[!ht]
\centering

\caption{Categorization by techniques type,applications domains and recommendations(continued).}
\label{tab:su}




\resizebox{18cm}{!}{
\begin{tabular}{|>{\large}c|>{\large}c|>{\large}c|>{\large}c|>{\large}c|>{\large}c|}
\hline
\textbf{Perspective}                                                                                            & \textbf{Reference}                                                     & \textbf{Technique}                                                                   & \textbf{Application}                & \textbf{Thing of interest}                                   & \textbf{Key contribution}                                                                                                                                                                                                          \\ \hline
\multirow{9}{*}{\textbf{\begin{tabular}[c]{@{}c@{}}Conventional \\ recommendations\\  techniques\end{tabular}}} & \begin{tabular}[c]{@{}c@{}}Tu et al.\\ \cite{tu2016context}\end{tabular}           & HB                                                                                   & smart marketing                    & \begin{tabular}[c]{@{}c@{}}service/\\ product\end{tabular}   & \begin{tabular}[c]{@{}c@{}}adapts HB to build recommender \\ system engine which makes\\  DS more attractive\end{tabular}                                                                                                          \\ \cline{2-6} 
                                                                                                                &  \cite{kim2018user}                                                           & HB                                                                                   & personal                            & \begin{tabular}[c]{@{}c@{}}service/\\ product\end{tabular}   & \begin{tabular}[c]{@{}c@{}}uses mobile edge environment \\ rather than center cloud to enhance\\  the quality of service recommendation\end{tabular}                                                                               \\ \cline{2-6} 
                                                                                                                & \begin{tabular}[c]{@{}c@{}}HamlAbadi et al.\\ \cite{hamlabadi2017framework}\end{tabular}    & HB                                                                                   & smart marketing                    & products                                                     & \begin{tabular}[c]{@{}c@{}}adapts a cognitive system to\\  build recommender  system in IoT\end{tabular}                                                                                                                           \\ \cline{2-6} 
                                                                                                                & \begin{tabular}[c]{@{}c@{}}Hwang et al.\\  \cite{hwang2016data}\end{tabular}        & KB                                                                                   & smart home                          & \begin{tabular}[c]{@{}c@{}}configuration\\ rule\end{tabular} & \begin{tabular}[c]{@{}c@{}}uses ontology and open\\  data to  build recommender system \\ for smart home\end{tabular}                                                                                                              \\ \cline{2-6} 
                                                                                                                & \begin{tabular}[c]{@{}c@{}}Kumar et al.\\   \cite{kumar2017smart}
                                                                                                                \end{tabular}        & KB                                                                                   & Personal                            & service                                                      & \begin{tabular}[c]{@{}c@{}}proposes a conceptual framework\\ for space service recommendations\end{tabular}                                                                                                                        \\ \cline{2-6} 
                                                                                                                & \begin{tabular}[c]{@{}c@{}}Varfolomeyev et al.\\ \cite{varfolomeyev2015smart}
                                                                                                                \end{tabular} & KB                                                                                   & smart tourism                       & service                                                      & \begin{tabular}[c]{@{}c@{}}implement recommender\\ system for historical tourism by\\  adapting ontology formats\end{tabular}                                                                                                      \\ \cline{2-6} 
                                                                                                                & \begin{tabular}[c]{@{}c@{}}Martino et al.\\ \cite{di2017fuzzy}\end{tabular}      & KB                                                                                   & personal                            & srevice                                                      & \begin{tabular}[c]{@{}c@{}}builds recommender system using\\ ontology   for  smart health field\end{tabular}                                                                                                                    \\ \cline{2-6} 
                                                                                                                & \begin{tabular}[c]{@{}c@{}}Ali et al.\\ \cite{ali2018type}\end{tabular}          & KB                                                                                   & personal                            & service                                                      & \begin{tabular}[c]{@{}c@{}}adapts fuzzy ontology to build recommender \\ system which recommends specific \\ food  for diabetes patients\end{tabular}                                                                              \\ \cline{2-6} 
                                                                                                                & \begin{tabular}[c]{@{}c@{}}Felfernig et al.\\ \cite{felfernig2017recommendation}\end{tabular}    & UB                                                                                   & smart home                          & software                                                     & \begin{tabular}[c]{@{}c@{}}apply the utility-based approach in\\ an AGILE project to recommend \\ technologies to a user\\ who employs the AGILE gateway\end{tabular}                                                              \\ \hline
\multicolumn{1}{|l|}{\multirow{8}{*}{\textbf{Context awareness}}}                                               & \begin{tabular}[c]{@{}c@{}}Salman et al.\\ \cite{salman2017model}\end{tabular}       & \begin{tabular}[c]{@{}c@{}}multi-type \\ context-aware\end{tabular}                  & personal                            & service                                                      & \begin{tabular}[c]{@{}c@{}}exploits context-aware to build \\ recommender system which provides multi \\ type of recommendation\end{tabular}                                                                                       \\ \cline{2-6} 
\multicolumn{1}{|l|}{}                                                                                          & \begin{tabular}[c]{@{}c@{}}Hassani et al.\\  \cite{hassani2018querying}\end{tabular}      & CSDL,CSM                                                                             & smart car park                      & srevice                                                      & \begin{tabular}[c]{@{}c@{}}propose a Context-as-a-Service (CoaaS) \\ recommender system\end{tabular}                                                                                                                               \\ \cline{2-6} 
\multicolumn{1}{|l|}{}                                                                                          & \begin{tabular}[c]{@{}c@{}}Yuan et al.\\ \cite{yuan2015and}\end{tabular}         & contextual information                                                               & social                              & service                                                      & \begin{tabular}[c]{@{}c@{}}exploit contextual information and\\ mobility of the user's tweets to \\ provide accurate recommendations.\end{tabular}                                                                                 \\ \cline{2-6} 
\multicolumn{1}{|l|}{}                                                                                          & \begin{tabular}[c]{@{}c@{}}Yavari et al.\\ \cite{yavari2016contextualised}\end{tabular}       & \begin{tabular}[c]{@{}c@{}}contextual filter,\\  contextual aggregation\end{tabular} & smart car park                      & service                                                      & \begin{tabular}[c]{@{}c@{}}exploits contextualization with IoT data\\  to provide accurate recommendations.\end{tabular}                                                                                                           \\ \cline{2-6} 
\multicolumn{1}{|l|}{}                                                                                          & \begin{tabular}[c]{@{}c@{}}Zhou et al.\\  \cite{zhou2017social}\end{tabular}         & HSCT algorithm                                                                       & personal                            & service                                                      & \begin{tabular}[c]{@{}c@{}}proposes a model which exploits \\ context-awareness to improve\\  recommendations accuracy\end{tabular}                                                                                                \\ \cline{2-6} 
\multicolumn{1}{|l|}{}                                                                                          & \begin{tabular}[c]{@{}c@{}}Kaur et al.\\ \cite{kaur2019context}\end{tabular}         & Context awareness                                                                    & smart kitchen                       & service                                                      & \begin{tabular}[c]{@{}c@{}}adapts context as sources\\ to build a recommender engine\\  for recipe recommendations\\ in a smart kitchen\end{tabular}                                                                               \\ \cline{2-6} 
\multicolumn{1}{|l|}{}                                                                                          & \begin{tabular}[c]{@{}c@{}}Casino et al.\\ \cite{casino2018smart}\end{tabular}       & Context awareness                                                                    & personal                            & service                                                      & \begin{tabular}[c]{@{}c@{}}proposes a recommender system\\ which provides \\ health recommendations\end{tabular}                                                                                                                   \\ \cline{2-6} 
\multicolumn{1}{|l|}{}                                                                                          & \begin{tabular}[c]{@{}c@{}}Hong et al.\\ \cite{hong2017social}\end{tabular}         & Context awareness                                                                    & \multicolumn{1}{l|}{smart tourism}  & service                                                      & \begin{tabular}[c]{@{}c@{}}proposes a recommender system\\  that provides social recommendations \\ for cultural heritage.\end{tabular}                                                                                            \\ \hline
\multirow{2}{*}{\textbf{Social IoT}}                                                                            & \begin{tabular}[c]{@{}c@{}}Saleem et al.\\ \cite{saleem2016exploitation}\end{tabular}       & SIoT                                                                                 & personal                            & service                                                      & \begin{tabular}[c]{@{}c@{}}exploits socialization between thing\\  to produce service recommendations\end{tabular}                                                                                                                 \\ \cline{2-6} 
                                                                                                                & \begin{tabular}[c]{@{}c@{}}Chen et al.\\
                                                                                                                \cite{chen2016scheme}\end{tabular}         & SIoT                                                                                 & unknown                             & device                                                       & \begin{tabular}[c]{@{}c@{}}provides recommendations about\\  trustworthy node\end{tabular} 
                                                                                                                 \\ \cline{2-6} 
                                                                                                                & \begin{tabular}[c]{@{}c@{}}Ren et al.\\
                                                                                                                \cite{ren2018recommender}\end{tabular}         & SIoT                                                                                 & social                             & service                                                       & \begin{tabular}[c]{@{}c@{}}the mobile IoT was exploited \\to build a recommender system for services and\\ social partners\end{tabular} 
                                                                                                            \\ \hline
\multicolumn{1}{|l|}{\multirow{3}{*}{\textbf{Multi-agent system}}}                                              & \begin{tabular}[c]{@{}c@{}}Forestiero \\ \cite{forestiero2017multi}\end{tabular}         & MAS                                                                                  & unknown                             & device                                                       & \begin{tabular}[c]{@{}c@{}}adapts a multi agent algorithm\\ to improve the recommendation's speed.\end{tabular}                                                                                                                \\ \cline{2-6} 
\multicolumn{1}{|l|}{}                                                                                          & \begin{tabular}[c]{@{}c@{}}Martino and Rossi\\  \cite{di2016architecture}
\end{tabular}   & MAS                                                                                  & \multicolumn{1}{l|}{smart car park} & service                                                      & \begin{tabular}[c]{@{}c@{}}builds a Mobility Recommender\\ System(MRS) for parking\end{tabular}                                                                                                                                    \\ \cline{2-6} 
\multicolumn{1}{|l|}{}                                                                                          & \begin{tabular}[c]{@{}c@{}}Twardowski and\\ Ryzko\cite{twardowski2015iot}\end{tabular}         & MAS                                                                                  & smart m-health                            & service                                                      & \begin{tabular}[c]{@{}c@{}}exploits the MAS to build\\ a recommender system which uses the data\\ of mobile devices and  some IoT \\devices surrounding them\\  to provide personalized recommendations\\ in real time.\end{tabular} \\ \hline
\end{tabular}}

  \end{table*}
\clearpage

\begin{table*}[!ht]
\centering
\caption{Categorization by techniques type,applications domains and recommendations(continued).}
\label{tab:su}
\resizebox{18cm}{!}{
\begin{tabular}{|>{\large}c|>{\large}c|>{\large}c|>{\large}c|>{\large}c|>{\large}c|}
\hline
\textbf{Perspective}                           & \textbf{Reference}                                                                                          & \textbf{Technique}                                                                                                                              & \textbf{Application}                  & \textbf{Thing of interest} & \textbf{Key contribution}                                                                                                                                                                                            \\ \hline
\textbf{\begin{tabular}[c]{@{}c@{}}Multi-agent \\ system\end{tabular}}                  & \begin{tabular}[c]{@{}c@{}}Jiménez\\Bravo et al.\\ \cite{jimenez2019multi}\end{tabular}                                     & MAS                                                                                                                                             & smart home                            & service                    & \begin{tabular}[c]{@{}c@{}}proposes a recommender system \\based on  the multi-agent \\technique to optimize electricity \\consumption and save \\cost in a smart home.\end{tabular}                                \\ \hline
\multirow{2}{*}{\textbf{\begin{tabular}[c]{@{}c@{}}Graph\\  Database\\  Model\end{tabular}}} & \begin{tabular}[c]{@{}c@{}}Palaiokrassas et al.\\   \cite{palaiokrassas2017iot}\end{tabular}                                     & Neo4j graph                                                                                                                                     & smart home                            & service                    & \begin{tabular}[c]{@{}c@{}}exploits a Neo4j \\graph database to \\address one of the \\main challenge in IoT,\\ namely, big data management.\end{tabular}                                                              \\ \cline{2-6} 
                                               & \begin{tabular}[c]{@{}c@{}}Noirie et al.\\ \cite{noirie2017towards}\end{tabular}                                            &    \begin{tabular}[c]{@{}c@{}}Typed\\  Attributed \\ Graphs\end{tabular}                                                                                                                          & unknown                               & service                    & \begin{tabular}[c]{@{}c@{}}exploits graph techniques to build a\\  recommender system which provides IoT \\ services recommendations to the\\ user based on their own IoT devices.\end{tabular}                      \\ \hline
\multirow{12}{*}{\textbf{\begin{tabular}[c]{@{}c@{}}Machine\\  Learning\end{tabular}}}      & \begin{tabular}[c]{@{}c@{}}Sewak et al. \\ \cite{sewak2016iot} \end{tabular}                                            & \begin{tabular}[c]{@{}c@{}}Distributed\\ Kalman Filters,\\  Distributed\\ Mini-Batch\\ SGD ,\\ and Distributed ALS\\ based classifier\end{tabular} & unknown                               & unknown                    & \begin{tabular}[c]{@{}c@{}}adapts some machine learning \\algorithms to build the Optimal State based \\Recommender(OSR)\end{tabular}                                                                                  \\ \cline{2-6} 
                                               & \begin{tabular}[c]{@{}c@{}}Guo et al.\\ \cite{guo2018mobile}\end{tabular}                                               & RBF,D-S                                                                                                                                         & \multicolumn{1}{l|}{smart marketing} & products                   & \begin{tabular}[c]{@{}c@{}}proposes a framework to build\\ an e-commerce recommender system that \\ exploited the multisources of information\end{tabular}                                                           \\ \cline{2-6} 
                                               & \begin{tabular}[c]{@{}c@{}}Massimo et al.\\ \cite{massimo2017learning}\end{tabular}                                           & IRL                                                                                                                                             & smart tourism                         & service                    & \begin{tabular}[c]{@{}c@{}}proposes inverse reinforcement learning (IRL)\\  to model user behaviour to improve\\ the quality of recommendations.\end{tabular}                                                        \\ \cline{2-6} 
                                               & \begin{tabular}[c]{@{}c@{}}Gutowski et al.\\ \cite{gutowski2017framework}\end{tabular}                                          & LinUCB                                                                                                                                          & personal                              & service                    & \begin{tabular}[c]{@{}c@{}}uses a reinforcement learning algorithm\\ to build framework that provides \\ context-aware recommendations in\\ a smart city.\end{tabular}                                               \\ \cline{2-6} 
                                               & \begin{tabular}[c]{@{}c@{}}Oyeleke et al.\\  \cite{oyeleke2018situ}\end{tabular}                                           & RL                                                                                                                                              & smart home                            & service                    & \begin{tabular}[c]{@{}c@{}}proposes a system too\\ recommend the correct sequence\\  of tasks for each activity\\ to ensure that the user can reach his goal.\end{tabular}                                           \\ \cline{2-6} 
                                               & \begin{tabular}[c]{@{}c@{}}Asthana et al. \\     \cite{asthana2017recommendation}\end{tabular} & \begin{tabular}[c]{@{}c@{}}a machine\\ learning classifier\end{tabular}                                                                         & smart m-health                           & device                     & \begin{tabular}[c]{@{}c@{}}exploits a machine learning classifier to build\\  a recommendation engine which provides\\  personalized wearable technologies recommendations\\  for proactive monitoring.\end{tabular} \\ \cline{2-6} 
                                               & \begin{tabular}[c]{@{}c@{}}Amoretti et al.\\  \cite{amoretti2017utravel}\end{tabular}       & \begin{tabular}[c]{@{}c@{}}K-Means,\\ UPR\end{tabular}                                                                                          & social                                & service                    & \begin{tabular}[c]{@{}c@{}}adapts K-Means algorithm in UTrave\\ recommender system application which clusters\\ user profiles\\ to recommend points of interest for the user.\end{tabular}                           \\ \cline{2-6} 
                                               & \begin{tabular}[c]{@{}c@{}}Rasch \cite{rasch2014unsupervised}  \\     \end{tabular}                                               & \begin{tabular}[c]{@{}c@{}}unsupervised \\ learning algorithm\end{tabular}                                                                      & smart home                            & service                    & \begin{tabular}[c]{@{}c@{}}adapts unsupervised learning to build\\  recommender system for smart home.\end{tabular}                                                                                                  \\ \cline{2-6} 
                        & \begin{tabular}[c]{@{}c@{}}Yoo and \\Chung \cite{yoo2018mining}  \end{tabular}                                                                  & decision tree                                                                                                                                   & smart m-health                              & service                    & \begin{tabular}[c]{@{}c@{}}uses a decision tree\\ to build a system which provides \\ lifecare recommendations.\end{tabular}                                                                                         \\ \cline{2-6} 
                                               & Valtolina \cite{valtolina2014user}                                                        & \begin{tabular}[c]{@{}c@{}}decision tree \\  and social network\end{tabular}                                                                    & unknown                               & service                    & \begin{tabular}[c]{@{}c@{}}adapts  both the decision tree algorithm \\ and social network to propose a \\ multi-level recommendation system\end{tabular}                                                             \\ \cline{2-6} 
                                               & Rizvi \cite{rizvi2018aspire}                                                               & AHP                                                                                                                                             & smart car park                        & service                    & \begin{tabular}[c]{@{}c@{}}adapts Analytic HierarchyProcess (AHP) to \\ build recommender system for \\ car parking recommendations.\end{tabular}                                                                    \\ \cline{2-6} 
                                               & Ayata \cite{ayata2018emotion}                                                             & \begin{tabular}[c]{@{}c@{}}random forest,\\kNN and\\ decision tree.\end{tabular}                                                                 & personal                              & service                    & \begin{tabular}[c]{@{}c@{}}adapt ML algorithms \\to design a\\ recommender system for\\ music recommendations with\\ more accuracy than traditional ones.\end{tabular}                                                  \\ \hline
\multirow{2}{*}{\textbf{\begin{tabular}[c]{@{}c@{}}Deep\\  Learning\end{tabular}}}     & Yong \cite{yong2018iot}                                                                    & 3D CNN                                                                                                                                          & smart m-health                              & service                    & \begin{tabular}[c]{@{}c@{}}utilizes deep learning technique to build\\ an intelligent system for fitness club.\end{tabular}                                                                                       \\ \cline{2-6} 
                                               & \begin{tabular}[c]{@{}c@{}}Hashemi \\and Kamps \\ \cite{hashemi2018exploiting}
                                               \end{tabular}  & MLP   
                                               & smart tourism                         & service                    & \begin{tabular}[c]{@{}c@{}}a deep neural MLP has been adapted \\ to improve effective system \\recommendations for smart museum\end{tabular}                                                                        \\ \hline
\multirow{3}{*}{\textbf{Other}}                &  
\begin{tabular}[c]{@{}c@{}}Cha \cite{cha2016role}
                                               \end{tabular}           
& geofencing                                                                                                                                      & smart tourism                         & service                    & \begin{tabular}[c]{@{}c@{}}a geofencing was exploited\\ to build a real-time recommender \\system with IoT platform\end{tabular}                                                                                    \\ \cline{2-6} 
                                               &
                                               
                        \begin{tabular}[c]{@{}c@{}}Choi \cite{choi2015recommendation}
                                               \end{tabular}

                                               & bandwagon effect                                                                                                                                & personal                              & item                       & \begin{tabular}[c]{@{}c@{}}the bandwagon effect is\\ adapted  to build recommender system\end{tabular}                                                                                                               \\ \cline{2-6} 
                                               & \begin{tabular}[c]{@{}c@{}}Kamal et al.\\    \cite{kamal2013autonomic}

                                               \end{tabular}           & \begin{tabular}[c]{@{}c@{}}Autonomic\\  Network \\ Management\end{tabular}                                                                                                                  & personal                              & service                    & \begin{tabular}[c]{@{}c@{}}builds a learning-based \\ Autonomic Network Management \\System to provide personalized\\ recommendations by\\ exploiting user emotions and environmental information\end{tabular}       \\ \hline
\end{tabular}}

  \end{table*}

\clearpage
\bibliographystyle{abbrv}
\bibliography{bb}
%

\end{document}